\documentclass[letterpaper, 11pt]{article} 
\usepackage[top=1in, bottom=1in, left=1in, right=1in]{geometry}

\usepackage[T1]{fontenc}
\usepackage[utf8]{inputenc}
\usepackage{amssymb}
\usepackage{amsmath,amssymb,amsthm}

\usepackage{epsfig}
\usepackage{arydshln}
\usepackage{subcaption}
\newcommand{\eeq}[1]{\label{#1}\end{equation}}

\newcommand{\CO}{{\cal O}}

\newcommand{\bpm}{\begin{pmatrix}}
\newcommand{\epm}{\end{pmatrix}}

\def\Be{{\bf e}}
\def\Bf{{\bf f}}
\def\Bg{{\bf g}}

\def\Bp{{\bf p}}
\def\Bq{{\bf q}}

\def\Bu{{\bf u}}
\def\Bv{{\bf v}}
\def\Bw{{\bf w}}
\def\Bx{{\bf x}}
\def\By{{\bf y}}

\def\BF{{\bf F}}
\def\BG{{\bf G}}

\def\BQ{{\bf Q}}
\def\BR{{\bf R}}

\def\BT{{\bf T}}
\def\BU{{\bf U}}

\def\BX{{\bf X}}

\usepackage{comment}
\newcommand{\nonum}{\nonumber \\}
\usepackage{xspace} 
\usepackage{color}
\newcommand\col{black}
\newcommand{\cb}{\color{\col}}
\newcommand{\cred}{\color{\col}}
\newcommand{\cn}{\color{black}}

\newcommand{\cp}{\color{\col}}
\newcommand{\crevvv}{\color{black}}

\begin{document}



\title{Limit analysis of strut nets}


\author{Ada Amendola$^1$, Antonio Fortunato$^1$, Fernando Fraternali$^1$, Ornella Mattei$^2$, \\Graeme W. Milton$^3$, Pierre Seppecher$^4$}

\date{\footnotesize{$^1$Department of Civil Engineering, University of Salerno, Fisciano (SA), Italy\\
	$^2$Department of Mathematics, San Francisco State University, CA, USA\\
         $^3$Department of Mathematics, University of Utah, Salt Lake City, UT, USA\\
            $^4$Institut de Mathematiques de Toulon, Universite de Toulon, France}}
        \normalsize
\maketitle
\begin{abstract}
Truss structures composed of members that work exclusively in tension or in compression appear in several problems of science and engineering, e.g., in the study of the resisting mechanisms of masonry structures, as well as in the design of spider web-inspired web structures.
This work generalizes previous results on the existence of cable webs that are able to support assigned sets of nodal forces under tension. We extend such a problem to the limit analysis of compression-only ‘strut nets’ subjected to fixed and variable nodal loads. \cb These systems provide discrete element models of masonry bodies, \cp which lie inside \cn the polygon/polyhedron with vertices at the points of application of the given forces (`underlying \cp masonry structures'). \cn It is assumed  that fixed nodal forces are combined with variable forces growing proportionally to a scalar multiplier (load multiplier), and that the supporting strut net is subjected to kinematic constraints at given nodal positions.
\end{abstract}





\section{Introduction}
\cn
Force networks are frequently employed to describe the mechanical response of web-like structures and discrete element models of continuous bodies. 
\cred This is, e.g., \cn the case of bodies with a unilateral mechanical response, \crevvv by which we mean \cn that they exhibit an elastic- or rigid-type behavior characterized by the development of no-tension or no-compression stress fields (admitting only negative or positive eigenvalues of the stress tensor, respectively) \cite{delpiero1998,miltonrspa2019,miltonSIAM2020}. 
The ‘master safe theorem’ formulated by Heyman in his seminal work on the statics of masonry structures \cite{heyman1995} states that a no-tension masonry arch is stable if any thrust line in equilibrium with the external loads can be found within the masonry. 
A rigorous mathematical framework for the limit analysis of no-tension materials has been developed by Del Piero \cite{delpiero1998} and \v{S}ilhav\'y \cite{silhavy2014}, among others, in appropriate function spaces.
The limit analysis of solids and structures is a methodology for detecting the collapse value of given loads, without studying the evolution of the equilibrium problem along the loading history \cite{Kamen1996}.
It was originally formulated for bodies made of perfectly plastic materials \cite{Drucker1952,Save1997}, and has been generalized by Del Piero  \cite{delpiero1998}  to bodies composed of \textit{normal linear elastic materials}. The latter include no-tension materials, which share with perfectly plastic materials the presence of a convex set of admissible stress fields, and a normality rule of inelastic deformations with respect to such a domain. In the case of no-tension materials, plastic strains are replaced by fracture strains, which may occur when the stress field lies on the boundary of the admissible stress domain (see also Giaquinta and Giusti \cite{Giaquinta1985}, Angelillo et al. \cite{AngelilloJomms2010}).
The results presented in \cite{delpiero1998,silhavy2014} prove that the maximum of all statically admissible multipliers  gives a \textit{collapse multiplier}, which is associated with a \textit{collapse mechanism} of the structure.
\cb
According to the notation introduced by \v{S}ilhav\'y in \cite{silhavy2014} for no-tension bodies, a loading condition is \textit{strongly compatible} if any square integrable, compression-only stress fields exist that are in equilibrium with the given body and surface forces (through the principle of virtual work). A loading condition is instead \textit{weakly compatible} if any compression-only stress fields exist, represented by measures in equilibrium with the given forces. 
The latter can exhibit singular parts that are concentrated on surfaces (3D case) or lines/curves (2D case). 
\newline
\cn
Structural engineering approaches to the limit analysis of masonry bodies have been carried out by different authors, through the search of compression-only truss structures (‘strut nets’), which satisfy the equilibrium equations with the given loads, and are fully contained within the body of the masonry 
\cite{odwyer1999,block2007,lucchesi2008,fortunato2010,defaveri2013,como2017}. \cb The forces carried by such structural networks can be regarded as singular stress fields statically admissible with the given loads, according to the analysis presented in \cite{silhavy2014}. \cn
In the 2D case, use has been made of polyhedral Airy stress functions to generate admissible force networks  \cite{Fraternali2002,fraternali2010,fortunato2013,milton2017}.
A mirrored, tension-only response is exhibited by spider orb webs loaded in the large displacement regime \cite{pugno,jiang}. The outstanding mechanical properties of such webs have encouraged several researchers to investigate the design and manufacturing of bioinspired spider-web-like membranes and metamaterials, over recent years  \cite{jin2017,miniaci2016,cao2021}.

The present work generalizes the results presented in \cite{miltonrspa2019,miltonSIAM2020} regarding the existence of cable webs under tension that can support a given set of nodal loads.
The generalization is multi-fold. First, we consider compression only force networks forming strut net models of masonry structures, which support a combination of fixed forces and variable forces applied at given nodes, with the latter growing proportionally to a scalar multiplier $\lambda$. 
\cb We assume that the members forming the supporting strut nets $\cal S$ do not undergo local buckling, due to a rigid response of the material in compression. Available literature results have 
shown that buckling effects are actually negligible in the presence of sufficiently small slenderness ratios (see \cite{Zani2009} and references therein).  \cn
A second generalization 
of the research presented in \cite{miltonrspa2019,miltonSIAM2020} 
consists of admitting that the movement of selected nodes of $\cal S$ can be restrained by external supports.
\cb The third and 
final generalization 
leads us to search for the extreme values of $\lambda$ in correspondence of which the applied loads can be supported by a compression-only structure. 
Under suitable regularization (or integrability) conditions of the singular stress fields associated with such \textit{limit load multipliers}, one can regard these quantities \crevvv 
as lower bounds on the magnitude of the collapse load multiplier \cn
for the underlying masonry structures \cite{delpiero1998, silhavy2014}.
\cn We focus our attention on the equilibrium problem of strut nets whose boundary is a convex domain in two- or three-dimensions.
We start presenting the strut net problem in section \ref{analytic}. Next, we address the
formulation of linear programming procedures for problems dealing with simply-connected domains (showing no holes or inclusions), as well as multiply-connected domains associated with the presence of polygonal `\textit{obstacles}'  (section \ref{linearprogramming}).
The effectiveness and accuracy of the given procedures are illustrated by examining a parade of numerical results dealing with benchmark examples of masonry structures. The analyzed structures are subject to fixed vertical forces, and horizontal forces that can grow proportionally to a load multiplier $\lambda$ (section \ref{numerics}) \cite{equivalent13}. Such a loading condition is aimed at reproducing the effects of wind forces or seismic loading, through an equivalent static analysis method \cite{equivalent13}.
We end by drawing concluding remarks and discussing directions for future work in section \ref{conclusions}.

\section{The strut net problem} \label{analytic}

\begin{figure*}[tbh]
    \centering	
    \includegraphics[scale=0.65]{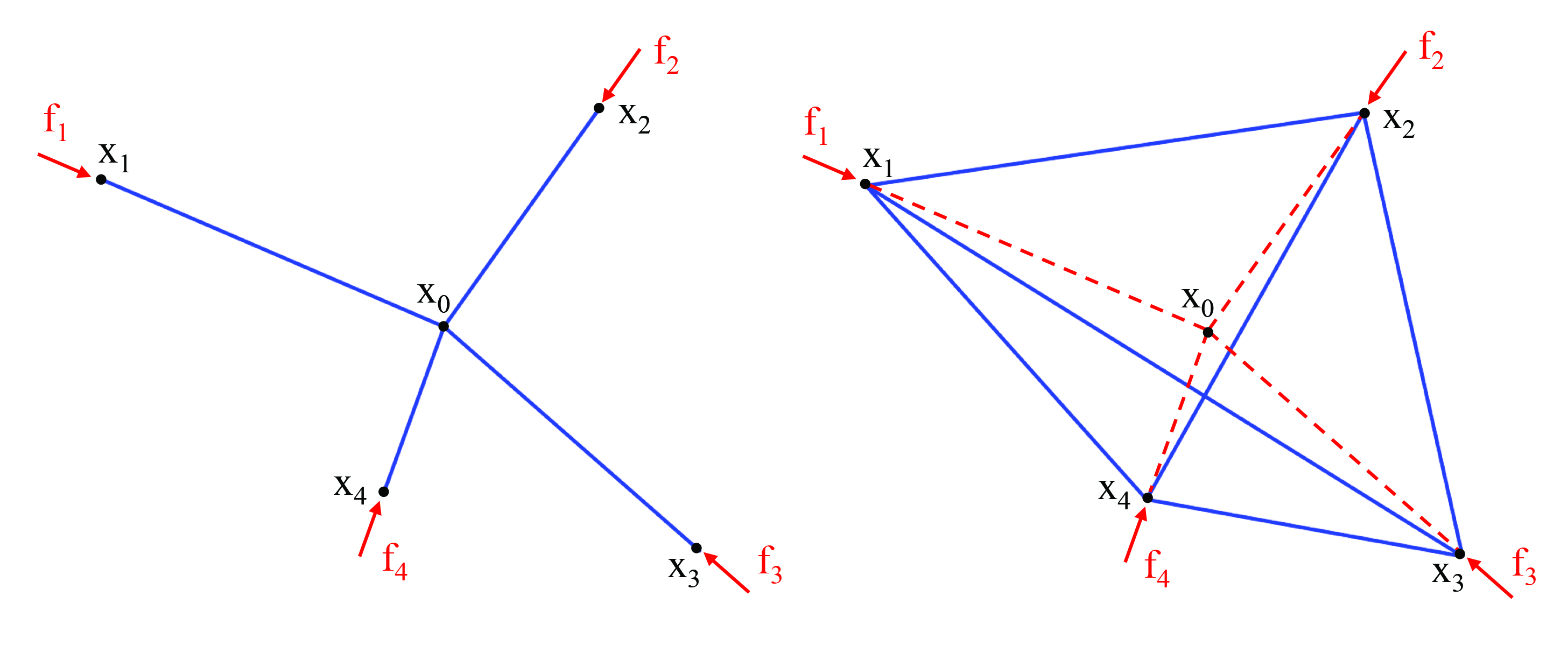} 
    \caption{Removal of one internal node at $\Bx_0$ that is linked to $n$  other nodes at $\Bx_1$, $\Bx_2$,..., $\Bx_N$. If $P_{j0}$ denotes the compression force in the strut from $\Bx_0$ to $\Bx_j$, then one can remove node $\Bx_0$ and replace it with a net in which the surrounding nodes are pairwise connected and have compression forces $P_{ij} =$  $\| \Bx_i-\Bx_j \| c_i c_j$ $($ $\sum_k$ $c_k$ $)^{-1}$, where $c_j$= $\|$ $\Bx_i$-$\Bx_j$ $\|$ $^{-1}$ $P_{0j}$ \cite{miltonSIAM2020}. Note that the examined structure is in three-dimensions and that the struts $\Bx_1-\Bx_3$ and $\Bx_2-\Bx_4$ do not touch each other.
    If a pair of these nodes were already connected, one should add this tension to the existing tension. Here, $\| \Bx \|$ notes the length of $\Bx$. Repeating this procedure allows one to remove all internal nodes.
    (Online version in color.)
    }
         \label{node_reduction}
\end{figure*}

Let us begin by reformulating the  ‘spider web problem’ recently studied in \cite{miltonrspa2019} to strut nets formed by members supporting compression-only forces: given a set of $N$ balanced forces $\Bf_1$, $\Bf_2$, ... , $\Bf_N$ at $N$  prescribed points $\Bx_1$, $\Bx_2$, ..., $\Bx_N$ in $d$-dimension ($d=2,3$), when does there exist a \cb \textit{supporting strut net} \cn $\cal S$ under compression, which connects the terminal points ${\BX}= ({\Bx}_1, {\Bx}_2, ..., {\Bx}_N)$,  and which carries axial forces in equilibrium with the given loading ${\BF}= ({\Bf}_1, {\Bf}_2, ..., {\Bf}_N)$?   The solution provided in Ref. \cite{miltonrspa2019} shows that it suffices to examine webs when no internal nodes exist, which means that one can easily address the query with linear programming \cite{miltonrspa2019}. Following the argument in \cite{miltonSIAM2020} and going one internal junction at a time, we may inductively replace the struts that are connected to this internal junction with a set of struts that pairwise join the nodes that are connected to the one they are replacing (figure \ref{node_reduction}). 
In a sense, this is similar to the "star-delta" transformation in resistor networks, though our result only applies to the stress and not to the elastic response of spring networks where one also monitors the displacement. A more sophisticated argument than this inductive one shows that the result holds true even if one starts with a continuum of struts \cite{miltonrspa2019}.

\begin{figure*}[tbh]
    \centering	
    \includegraphics[scale=0.625]{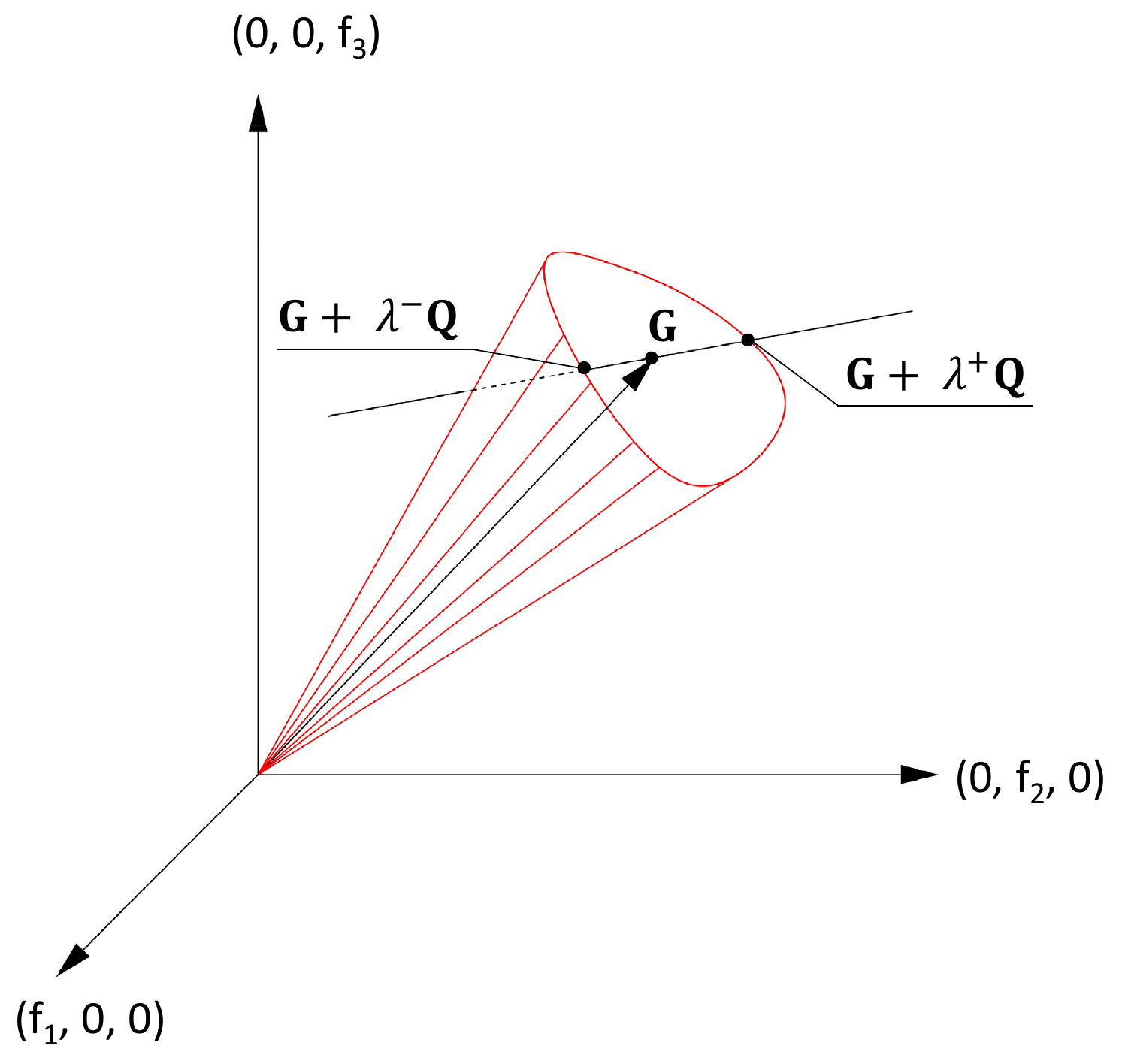} 
    \caption{Loading path and cone of the admissible loadings for $d=2$ and $N=3$. The figure is schematic as it is embedded in a 6-th dimensional space, and due to equilibrium of forces the cone lies in a hyperplane, \crevvv but otherwise has an arbitrarily shaped convex cross section.  \cn
    (Online version in color.)
    }
         \label{cone_figure}
\end{figure*}

\cb The limit-analysis formulation of the above problem is as follows. 
Let ${\Bu}_j$ denote the displacement vector of $\Bx_j$ \crevvv, \cn and let ${\BU}= ({\Bu}_1, {\Bu}_2, ..., {\Bu}_N)$  denote the global vector collecting all the nodal displacements of the given points. We introduce the following set of nodal displacements
\begin{equation}
\label{calA}
{\cal A} \ = \ 
\{
\BU \ | \ (\Bu_i - \Bu_j) \cdot (\Bx_i - \Bx_j) \le 0 , \ 1 \le i < j \le N 
\}
\end{equation}
and examine the \textit{loading path} $\BF(\lambda)=\BG + \lambda \BQ$, where $\BG=({\Bg}_1, {\Bg}_2, ..., {\Bg}_N)$ is a vector of fixed nodal forces; $\BQ=({\Bq}_1, {\Bq}_2, ..., {\Bq}_N)$ is 
the vector of \textit{proportional loads}
that defines the ‘shape’ of the  examined incremental loading process in $(\mathbb{R}^d)^N$, with origin at $\BG$; and $\lambda$ is a scalar quantity referred to as the \textit{loading multiplier}. We say that $\lambda$ is \textit{statically admissible} if there exists a supporting strut net for $\BF(\lambda)$.  Assuming that a supporting strut net for the fixed forces $\BG$ exists, it is easy to prove that the admissible loading multipliers are such that it results 
\begin{equation}
\label{sup}
\sup_{\BU \in \cal A} (\BG \ + \ \lambda \BQ) \ \cdot \BU \ \le 0,
\end{equation}
and that these multipliers form a suitable interval $(\lambda^-, \lambda^+)$ of  $\mathbb{R}$. Such a conclusion is a straightforward generalization of Theorem 1.1 of \cite{miltonrspa2019}. Consider indeed that the set of all the admissible loadings $\BF$ that are applied to $\BX$ \crevvv forms \cn a convex cone $\cal C$ in $(\mathbb{R}^d)^N$ \cite{miltonrspa2019}. \crevvv Intersection \cn points of $\BF(\lambda)$ with $\cal C$ give the \textit{limit load multipliers} $\lambda^-$ and $\lambda^+$ (see, e.g., the example in Fig. \ref{cone_figure}).

As we have already noted, if one wants a strut network under compression that is as robust as possible to different loadings, then all of this network's terminals should be connected pairwise. Internal nodes are not needed. An algorithmic approach to such a problem is presented in section \ref{linearprogramming}(a), assuming that selected nodes may be subject to kinematic constraints.
One observes, however, that internal nodes are needed if one wants all the struts to be contained in a desired region, which avoids selected ‘obstacles’. We will discuss such a problem in two-dimensions in section \ref{linearprogramming}(b), making use of a stress function formulation of the equilibrium problem. 
\cn

\newpage

\section{Linear programming procedures} \label{linearprogramming}

\cb

The present section presents two linear programming (LP) procedures for the limit analysis problem of strut nets in two- and three-dimensions. The first LP algorithm deals with a strut net that covers a compact domain in two- or three-dimensions without holes and/or inclusions. The second procedure instead refers to a 2D strut net that supports external forces at the boundary of a convex polygon and avoids an arbitrary number of polygonal obstacles \cite{obstacle22}. 

\cn

\subsection{Compact domains} \label{linearprogramming3D}

Let us consider a strut net which consists of a \cred \textit{complete net} \cn formed by all the pairwise connections of a set of $N$ nodes. \cb Such nodes are at the vertices of a compact and convex polygon/polyhedron. \cn
We write the equilibrium problem of $\cal S$ into the following matrix form
\begin{equation}
\label{equilibrium}
\bold{A} \bold{P} \ = \ \bold{G}  \ + \ \lambda \ \bold{Q}
\end{equation}
where $\bold{A}$ is the equilibrium matrix of the \cred complete net \cn (refer, e.g., to \cite{MRCtensegrity,SMStensegrity} for the expression of such a matrix), and  $\bold{P}=(\Bp_1,\Bp_2,\ldots,\Bp_n)$ is the vector of the axial forces carried by the members of $\cal S$. We let $n$ denote the total number of applied external forces, and let $m$ denote the total number of the pairwise connections of the points of applications of such forces.
We account for the presence of constraints setting selected nodal displacements of $\cal S$ to zero, by dropping the equilibrium equations associated with such degrees of freedom into problem \eqref{equilibrium} \cred (see, e.g., \cite{gill}, page 75). \cn
Assuming that \eqref{equilibrium} admits solutions for $\lambda=0$; post-multiplying both members of  such an equation by $\bold{Q}$; and solving for $\lambda$, we obtain
\begin{equation}
\label{lamba}
\lambda \ = \ \frac{1}{Q^2} \bold{A} \bold{P} \cdot \bold{Q} \ - \ \frac{1}{Q^2} \bold{Q} \cdot \bold{G}
\end{equation}
with $Q^2=\bold{Q} \cdot \bold{Q}$, ($\cdot$) denoting the symbol of the scalar product between vectors.

We search for the first \cred \textit{limit load multiplier} \cn $\lambda ^+$ of the proportional loads $\bold{F}$ by solving the following LP problem 
\begin{align}  
\underset{\bold{P}}{\text{\rm maximize}}  
	& \quad  {\lambda} \ = \ \bold{\bar{C}} \cdot \bold{P} 
	\ - \ \frac{1}{Q^2} \bold{Q} \cdot \bold{G}
	\nonumber \\
\text{\rm subject to}  
& \quad \left\{\begin{array}{l}
	\bold{\bar{A}} \bold{P} = \bold{\bar {Q}}  \\
	\bold{l_b} \leq  \bold{P} \leq  \bold{u_b} 
	\end{array} \right., 
\label{eq:lin_prog}
\end{align}  
Here, we have set 
\begin{equation}
\label{Abar}
\bold{\bar{C}} \ = \ \frac{1}{F^2} \bold{{A}}^T \bold{{F}};
\ \ \ \ 
\bold{\bar{A}} \ = \ \bold{{A}} \ - \ \frac{\bold{Q} \cdot \bold{A}}{Q^2} \ \bold{{Q}};
\ \ \ \ 
\bold{\bar{Q}} \ = \ \bold{{G}} \ - \ \frac{\bold{Q} \cdot \bold{G}}{Q^2} \ \bold{{Q}}
\end{equation}
where $\bold{{A}}^T$ denotes the transpose of $\bold{{A}}$.
In equation \eqref{eq:lin_prog}, due to the assumption of a rigid-no-tension behavior of the material with infinite compression strength,  $\bold{l_b}$ is a vector with all $-\infty$ entries, while $\bold{u_b}$ is a vector with all $0$ entries.
By \cred turning problem \eqref{eq:lin_prog} into a minimization problem, \cn we next obtain the second collapse multiplier $\lambda^{-}$. 
The vector of axial forces $\bold{P}$ that corresponds to prescribing $\lambda=\lambda^{+}$ (or $\lambda=\lambda^{-}$) in the  \cred \textit{limit load strut net} \cn is given by the solution of the LP problem \eqref{eq:lin_prog}.

\cb

\subsection{Obstacle problem in 2D} \label{linearprogramming2D}

We hereafter present a generalization of a result given in Ref. \cite{obstacle22}, which is aimed at handling limit analysis problems.
We search for a planar strut net, which supports a system of $N$ external forces applied at the vertices of a convex polygon $\Omega$, and avoids a number $s$ of polygonal obstacles $\CO_1$, $\CO_2$, \dots $\CO_s$. The generic obstacle $\CO_q$ is the convex hull of $N(q)$ points $(\By^q_1,\By^q_2,\dots,\By^q_{N(q)})$ and represents a region not accessible to the strut net (e.g., the region underneath an arch; a hole to be drilled in a masonry structure; an inclusion in the design domain formed by a non-reactive material or a void; etc.). As in the previous case, we assume that the external forces may be either active or reactive. The active forces are associated with the set ${\mathcal I}\subset \{1,\ldots,N\}$, while the reactive forces are associated with the complementary subset ${\mathcal K}:= \{1,\ldots,N\}\setminus {\mathcal I}$.

A straightforward generalization of Theorem 2 of Ref. \cite{obstacle22} leads us to identify the load multiplier of the {loading path} $\BF(\lambda)=\BG + \lambda \BQ$ with the objective function of the following linear programming problem
\begin{align}
&  \BR_\perp(\Bw_{i+1}-\Bw_{i})=-(\Bg_i + \lambda \Bq_i),
\ \ \ \text{for $i\in {\mathcal K}$,} \nonum 
&\Bw_1=0,\quad d_1=0,  \nonum 
&\Bw_{i+1}\cdot \Bx_i+d_{i+1}=\Bw_{i}\cdot \Bx_i+d_i,
\ \ \ \text{for $1\leq i\leq n$,} \nonum 
&\Bw_{j}\cdot \Bx_i+d_{j}\geq \Bw_{i}\cdot \Bx_i+d_i,
\ \ \ \text{for $i\not= j$ in $\{1,\dots,N\}$,} \nonum
& \Bv^{(q)}\cdot \Bx_i +c^{(q)}  \geq  \Bw_{i}\cdot \Bx_i+d_i,
\ \ \ \text{for $1\leq i\leq N$, $1\leq q\leq s$,} \nonum 
&\Bw_{i}\cdot \By_p^{(q)}+d_i \geq \Bv^{(q)}\cdot \By_p^{(q)}+c^{(q)},
\ \ \ \text{for $1\leq i\leq N$, $1\leq q\leq s$, $1\leq p\leq N(q)$,} \nonum 
&\Bv^{(r)}\cdot \By_p^{(q)}+c^{(r)} \geq \Bv^{(q)}\cdot \By_p^{(q)}+c^{(q)},
  \ \ \ \text{for $q\not= r$ in $\{1,\dots,s\}$, $1\leq p\leq N(q)$.}
    \label{LP2D}
\end{align}

\noindent Here, $ \BR_\perp=\bpm 0 & -1 \\ 1 & 0 \epm $ is the rotation matrix by an angle $\Pi/2$; $\Bw_{i}$ and $d_{i}$ respectively are a vector and a scalar that identify the tangent plane to the \textit{strut net function} $\phi$ along the generic segment of the boundary of $\Omega$; $\Bv_{i}$ and $c_{i}$ represent a vector and a scalar that define the value of $\phi$ over the generic obstacle \cite{obstacle22}. 
The strut net function $\phi$ is a polyhedral Airy stress function that generates the axial forces $\Bp_i$ carried by the limit load strut net.
Upon casting problem \eqref{LP2D} in the form of a maximization or a minimization problem, one computes $\lambda^{+}$ and $\lambda^{-}$, respectively.
\cn

\section{Numerical results} \label{numerics}

We hereafter present a parade of numerical applications of the LP procedures presented in the previous section, which deal with strut net models of masonry walls and arches.  
We analyze structures subjected to fixed vertical forces and variable horizontal forces, with the latter growing through a scalar multiplier $\lambda$ from a base value. 
In most cases, the base value of the horizontal forces is set equal to the resultant of the vertical forces, so that $\lambda$ can be identified with the spectral acceleration that describes an equivalent, static seismic loading condition of the structure \cite{equivalent13}. 
For the sake of simplicity, we will restrict our attention to loading paths of the form: $\BF(\lambda)=\BG + \lambda \BQ$ with $\lambda \ge 0$ \cite{delpiero1998,silhavy2014}, by searching for the limit load multiplier $\lambda_{lim}=\lambda^{+}$.
It is easily verified that it results $\lambda^{-}=0$ for all the examples that follow.

\subsubsection{Shear walls} \label{shearwalls}

\begin{figure*}[tbh]
    \centering	
    \includegraphics[scale=0.470]{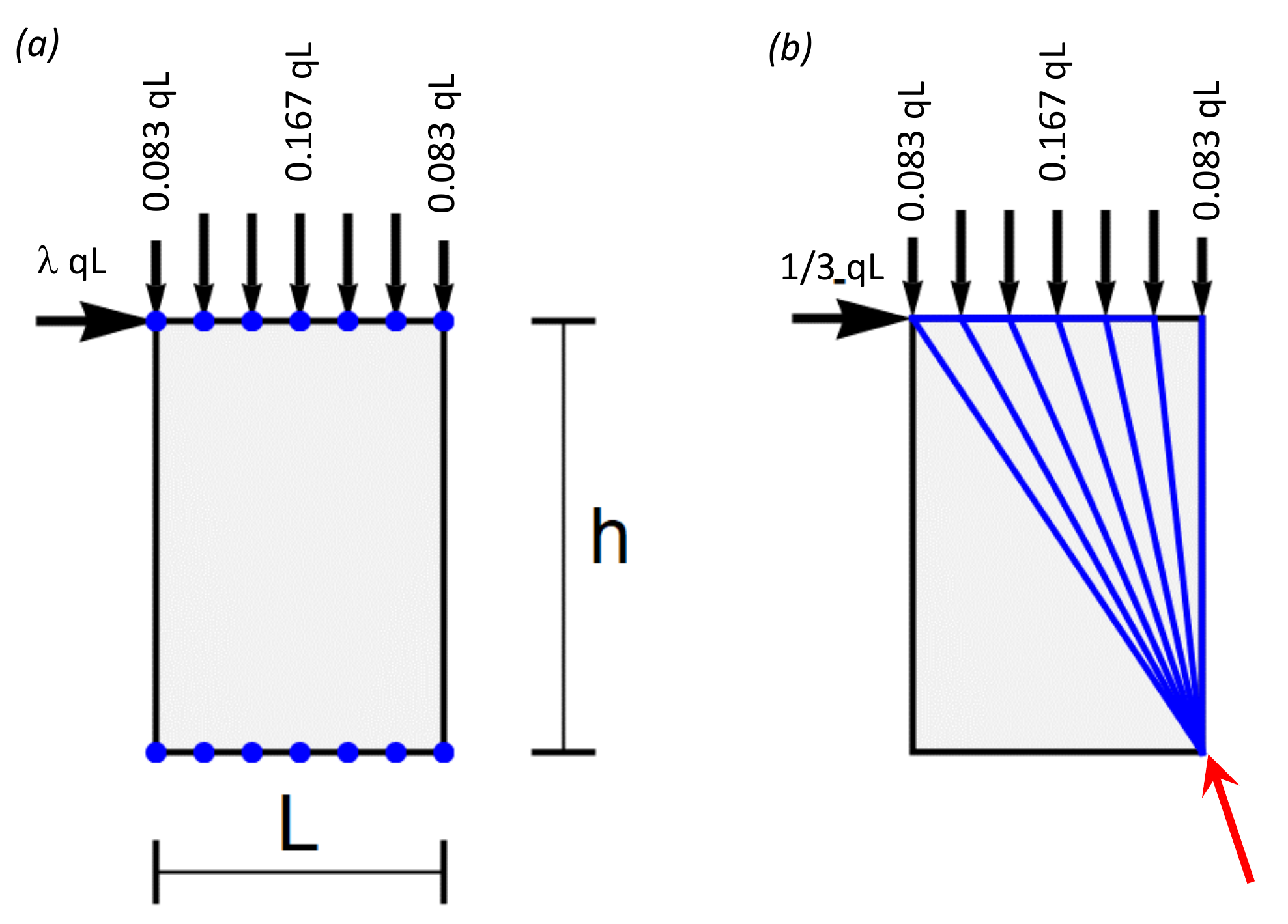} \\
    \includegraphics[scale=0.470]{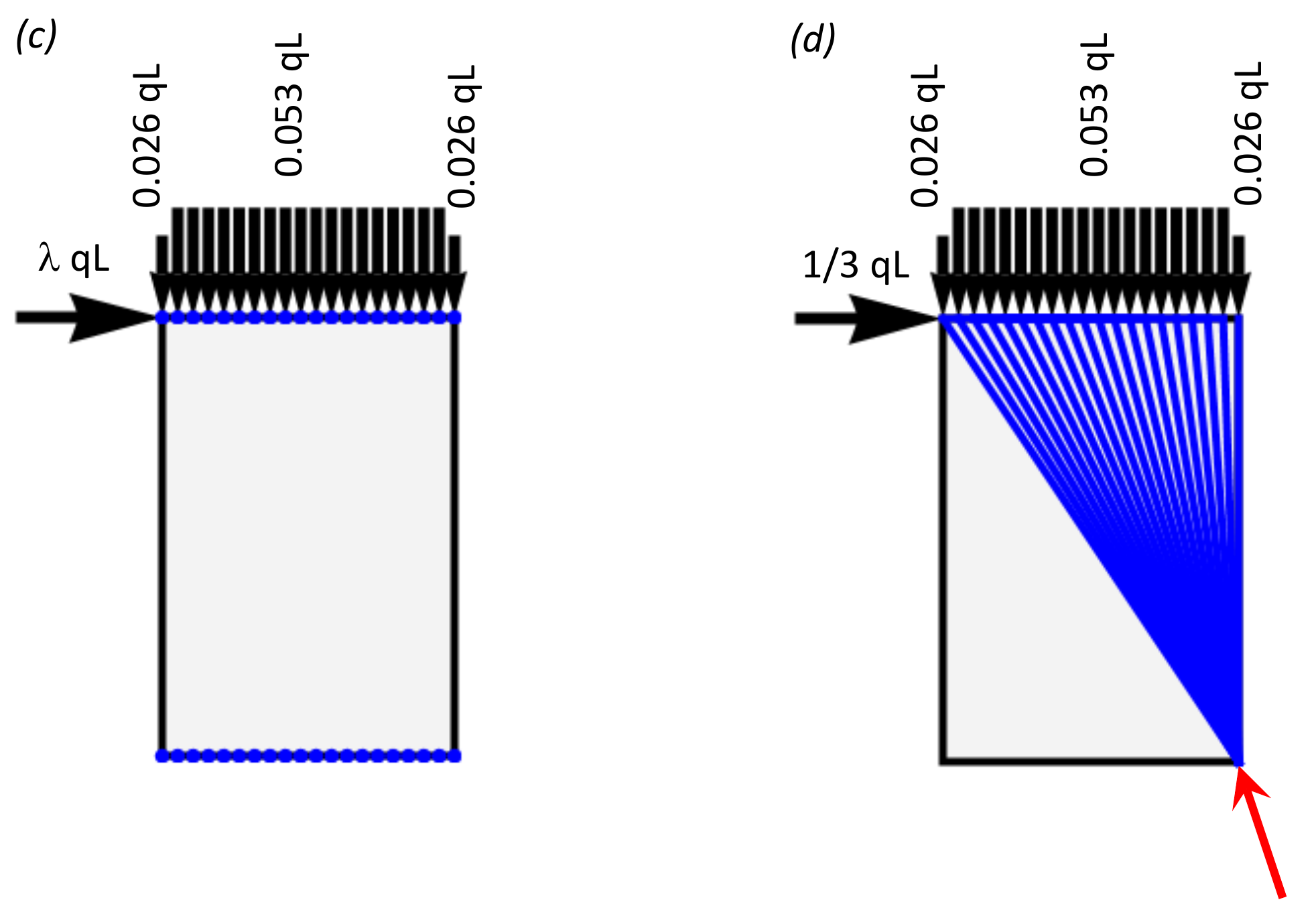} \\
    \caption{Limit analysis of strut net models of a masonry wall loaded by a uniformly distributed  vertical load and a concentrated horizontal force applied to the top edge of the wall.
    (\textit{a}) The complete net of a coarse model is formed by the pairwise connections of seven nodes placed on the top and bottom edges of the structure.
    (\textit{b}) The corresponding limit load multiplier $\lambda_{lim}$ is equal to $1/3$, as in the  limit analysis problem of the wall modeled as a continuum medium \cite{fortunato2013}. The examined load condition is supported by struts placed above a diagonal of the panel.
    (\textit{c}) A more refined model of the wall is obtained by examining the pairwise connections of groups of twenty nodes placed over the top and bottom edges of the wall.
    (\textit{d}) Also in this case, the LP procedure predicts $\lambda_{lim}=1/3$, and a limit load strut net formed by top-edge struts and rays converging to the bottom-right corner of the wall. In both the coarse and the refined models, the base reaction of the bottom-right node displays a vertical component equal to qL and a horizontal component equal to 1/3 qL. (The forces shown in figure  are not to scale. Online version in color.)}
    \label{panel_figure}
\end{figure*}

Let us consider first a rectangular wall serving as a supporting element of a masonry building, which exhibits horizontal span $L$ and height $h$.
With the aim of replicating the example presented in section 3.2 of Ref. \cite{fortunato2013}, we assume $h/L = 2/3$, and apply a distributed load $q$ to the top edge of the middle-plane of the wall, and a horizontal (shear) force $F=\lambda qL$ to the right corner of the same edge (figure \ref{panel_figure}). 

A first simulation examines the lumping of the vertical load $q$ at seven nodes, and introduces a mirrored discretization of the bottom edge of the middle-plane of the wall, which is formed by fixedly constrained nodes (complete web formed by the 91 pairwise connections of the 14 nodes shown in figure \ref{panel_figure}\textit{a}; limit load strut net formed by the 12 ‘active’ connections shown in figure \ref{panel_figure}\textit{b}). 
The LP problem \eqref{eq:lin_prog} returns the limit load multiplier \cred $\lambda_{lim}=\lambda^+=1/3$, \cn which coincides with the collapse  multiplier of the continuum model of the wall examined in \cite{fortunato2013}. The latter is associated with a rigid-body rocking mechanism of the wall about the bottom-right corner (see also \cite{como2017}). The \cred strut net corresponding to the limit load \cn of $\cal S$ is illustrated in figure \ref{panel_figure}\textit{b}. 
\cb It clearly results in $\lambda^-=0$, since it is easily proven that a horizontal force applied to the top-left corner of the wall and pointing leftward cannot be supported by any no-tension force network connecting the given points. On the other hand, we would find $\lambda^-=-1/3$  and $\lambda^+=0$ for the loading condition with the same vertical forces and a horizontal force applied to the top-right corner of the wall, which points rightward. \cn
A more refined discretization of the wall under examination is obtained by lumping the distributed load $q$ at twenty nodes of the top edge, and introducing and equal number of fixedly constrained nodes at the bottom edge (complete web formed by the 780 pairwise connections of the 40 nodes shown in figure \ref{panel_figure}\textit{c}). Also in this case the LP procedure returns $\lambda_{lim}=1/3$, and one notes that the limit load strut net is formed by a set of 38 active connections that include 18 top horizontal struts and 20 ‘rays’ converging to the bottom-right corner of the wall (figure \ref{panel_figure}\textit{d}).

\begin{figure*}[tbh]
    \centering	
    \includegraphics[scale=0.40]{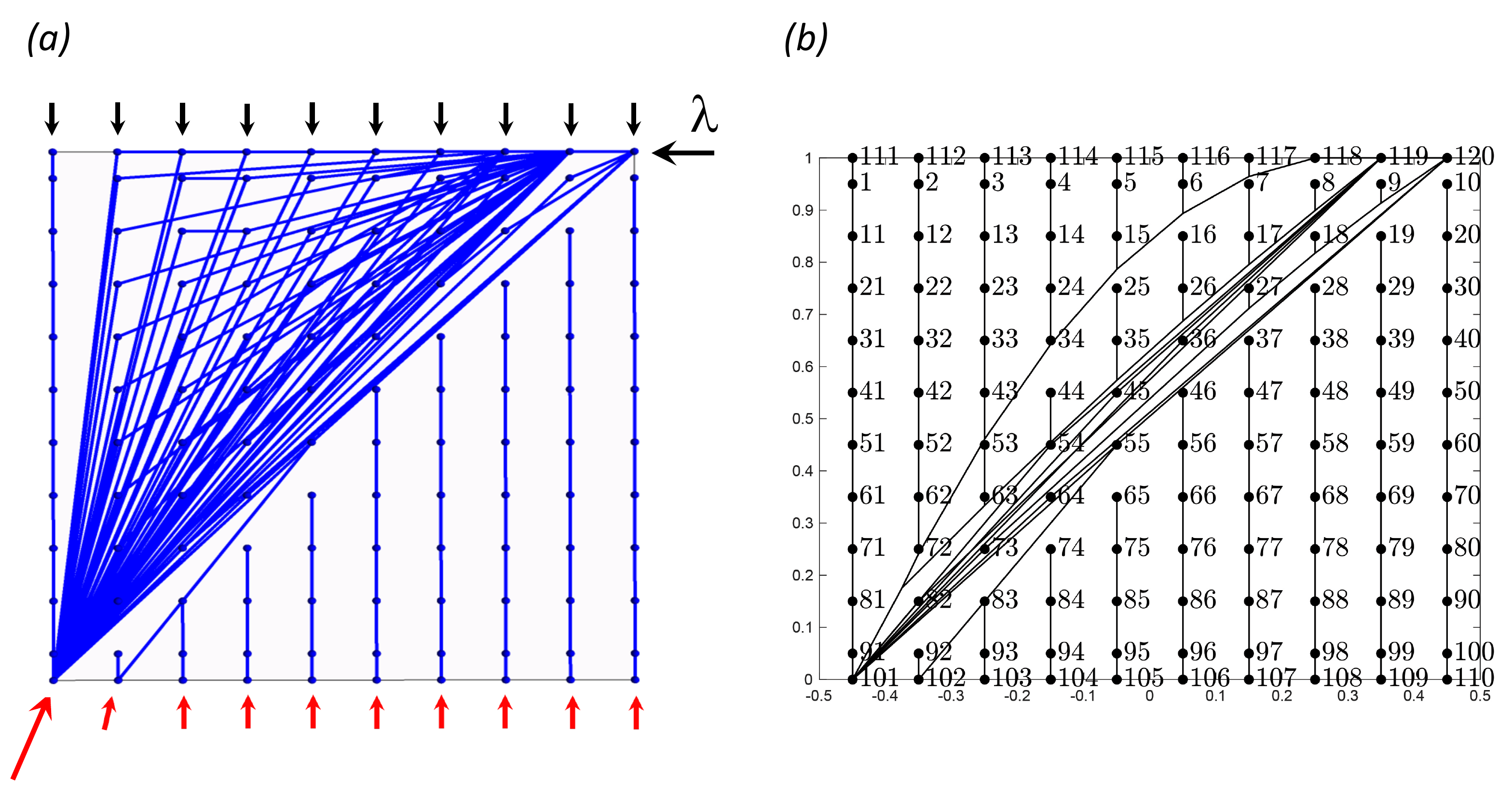} \\
    \includegraphics[scale=0.38]{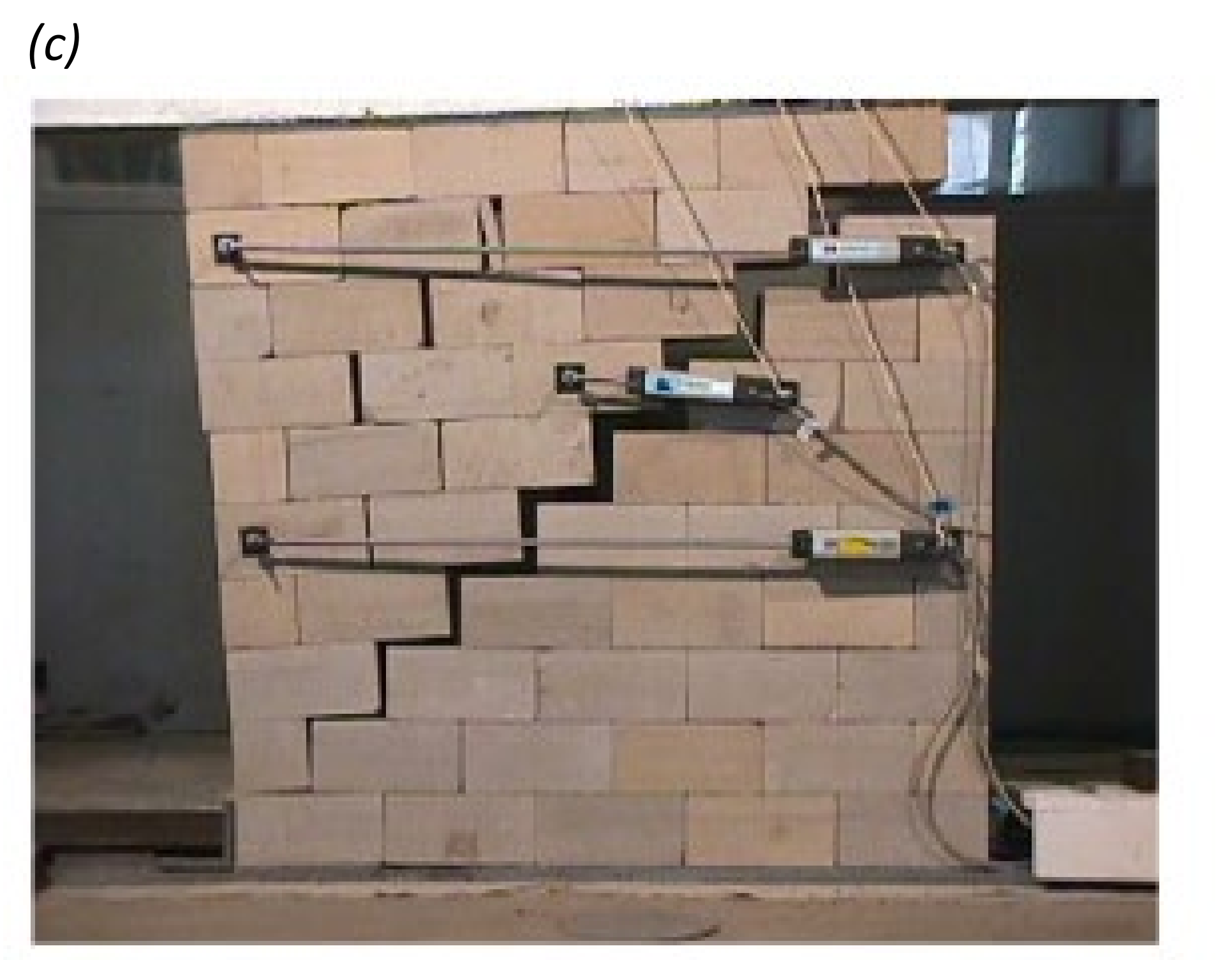} \\
    \caption{(\textit{a}) Strut net modeling of a dry-jointed masonry wall under the action of self-weight and a progressively increasing shear force. A grid of 10$\times$12 nodes is loaded by vertical forces with magnitude 0.05 kN applied at the inner nodes (not shown in figure); vertical forces with magnitude of 3 kN applied on the top nodes and a horizontal force with magnitude $\lambda$ applied to the right-top corner. The LP procedure of section \ref{linearprogramming3D} predicts a limit load $\lambda_{lim}=14.417$ kN associated to a strut net formed by oblique and vertical struts, with the oblique struts mainly placed above the main diagonal of the wall. The bottom-left reaction force has a vertical component of 32.47 kN and a horizontal component of 14.13 kN. The neighbor node applies a vertical reaction of 0.077 kN and a horizontal reaction of 0.018 kN. The remaining nodes of the bottom edge of the wall apply vertical reactions growing from 0.1 kN to 0.5 kN.
    (\textit{b}) Simplification of the strut net of panel (a) making use of the loop reduction procedure presented in \cite{miltonrspa2019}.
    (\textit{c}) The crack pattern experimentally observed by Lorenc\c{o}  et al. \cite{lorenco05} on a physical sample is consistent with the connectivity of the limit load strut net shown in panel (b), reproduced with permission from \cite{orduna17}. 
     (The forces shown in the figure are not to scale. Online version in color.)}
    \label{orduna_shear_figure}
\end{figure*}

We now focus our attention on a different shear wall problem, which is inspired by the experimental study on dry stone walls reported in \cite{orduna17} (Fig. \ref{orduna_shear_figure}).
We analyze a dry-joint stone wall with 1,000 mm width, 1,000 mm height and 200 mm thickness, which is subject to self-weight (25 kN/m$^3$ specific weight), a supplementary vertical load of 30 kN is applied to the top edge of the wall, and a horizontal load acting on the same edge. Such a shear load increases until it produces the collapse of the structure.   
A strut net model of the current example is obtained by introducing a 10$\times$12 grid of nodes over the mid-section.
The examined grid is formed by a first row of ten fixedly constrained nodes at the bottom edge of the wall; a second row of ten nodes at 50 mm height from the bottom; nine subsequent rows of ten nodes with 100 mm vertical spacing; and a row of ten nodes placed along the top edge of the wall. The complete web is formed by the 7,140 pairwise connections of these nodes. The self-weight is lumped at all the nodes of the above grid (0.05 kN in each node), while the supplementary vertical load is lumped at the ten nodes of the top edge (3 kN in each node). Finally, the horizontal force $F=\lambda$ is applied to the top-left corner of the same edge. The LP procedure of section \ref{linearprogramming3D} predicts a limit load multiplier $\lambda_{lim}=14.417$ kN, which is supported by the strut net formed by 157 active connections shown in figure \ref{orduna_shear_figure}\textit{a}.
On employing the loop reduction procedure that is detailed in \cite{miltonrspa2019} (see also section \ref{analytic}), we were able to reduce the strut net of figure \ref{orduna_shear_figure}\textit{a} to the simplified net that is illustrated in figure \ref{orduna_shear_figure}\textit{b} (the step-by-step loop reduction procedure is illustrated by  Movie S1 in the supplementary materials). 
Experimental tests carried out by Lorenc\c{o}  et al. \cite{lorenco05} on physical samples highlighted the onset of sliding displacements between the blocks under a value of the horizontal force approximately equal to 14-15 kN \cite{orduna17} (the crack pattern reported in \cite{lorenco05} is shown in figure \ref{orduna_shear_figure}\textit{c}).
It is worth noting that the no-tension model of masonry structures assumes that sliding failure cannot occur within the material \cite{delpiero1998,heyman1995}. It is therefore reasonable that the strut net model analyzed in this study is able to reproduce the experimental results reported in \cite{lorenco05,orduna17} as long as sliding effects are negligible, due to the gravity-induced compaction of the blocks \cite{heyman1995}.
The presence of struts above the main diagonal of the wall in figure \ref{orduna_shear_figure}\textit{b} is consistent with the crack pattern observed during the experimental tests presented in \cite{lorenco05,orduna17} (compare figures \ref{orduna_shear_figure}\textit{b,c}).
It is easily recognized that the system of vertical forces corresponding to setting  $\lambda=0$ is statically admissible for each of the shear walls examined in the present section. Such a loading condition is indeed supported by strut nets with vertical members passing through the points of application of the active and reactive forces present on the opposite edges of the wall.

\subsection{Walls with openings} \label{walls}

\begin{figure*}[h!]
	\centering
\begin{subfigure}{1.00\textwidth}
		\centering
		 \ \ \ 	\ \ \ \ \ 
		{(\textit{a})}: $n_a=21$, $\lambda=0$
          \ \ \ \ \ \ \
		{(\textit{b})}: $n_a=21$, $\lambda_{lim}=0.35833$
	\end{subfigure}
   \includegraphics[scale=0.525]{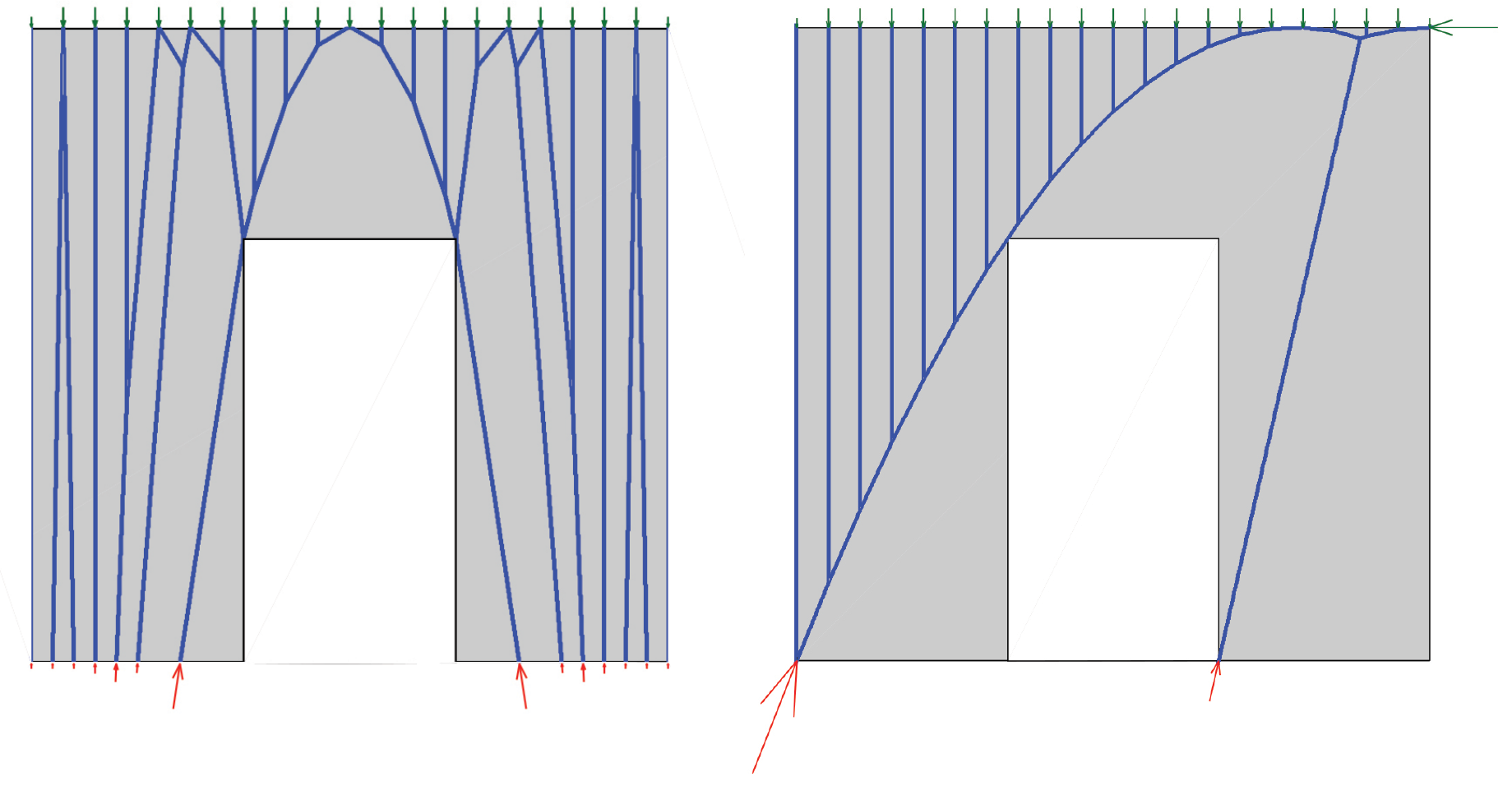} \\
\begin{subfigure}{1.00\textwidth}
		\centering
		 \ \ \ 	\ \ \ \ \ 
		{(\textit{c})}: $n_a=81$, $\lambda=0$
          \ \ \ \ \ \ \
		{(\textit{d})}: $n_a=81$, $\lambda_{lim}=0.35906$
	\end{subfigure}
   \includegraphics[scale=0.525]{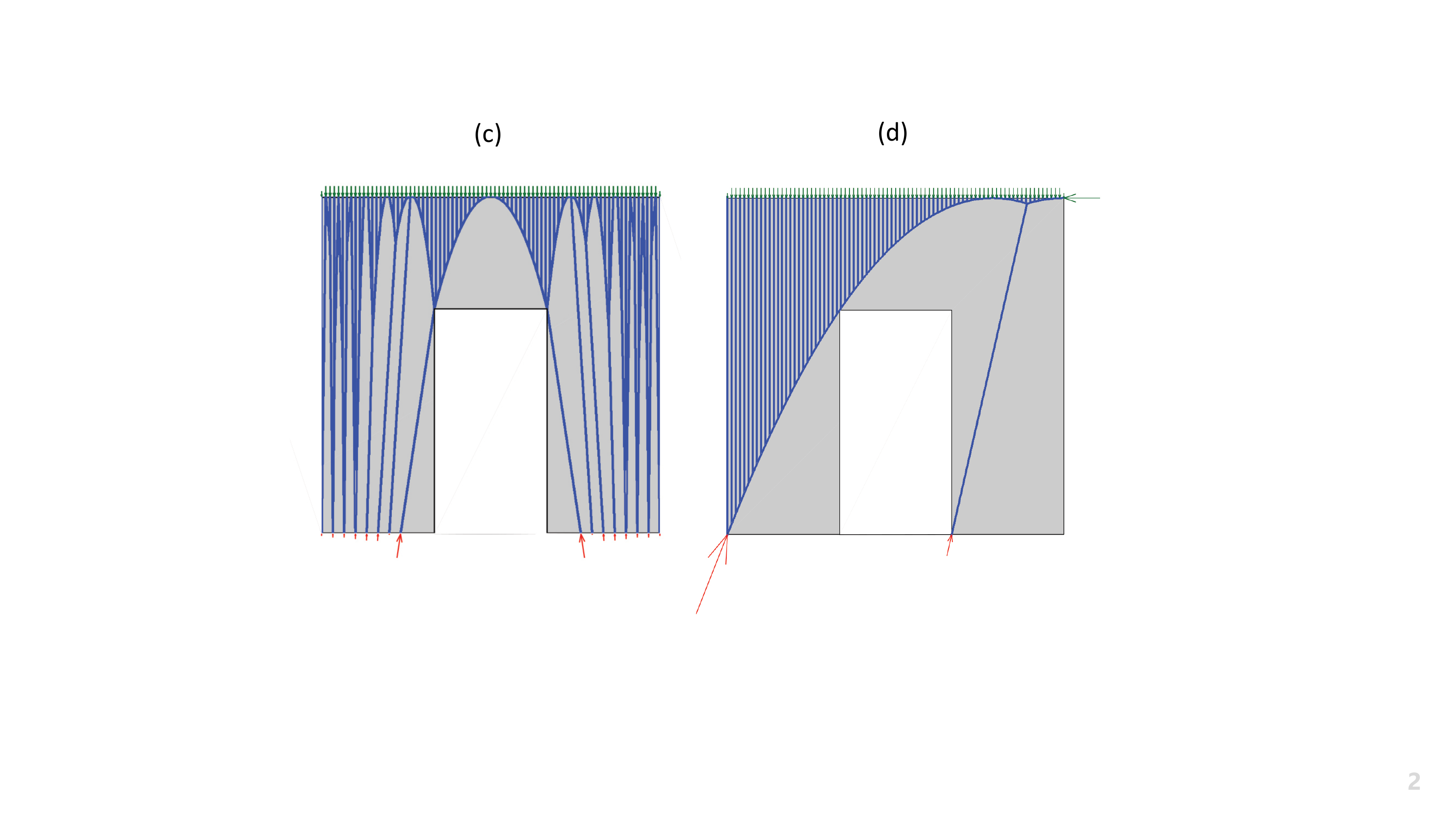} \\
\begin{subfigure}{1.00\textwidth}
		\centering
		 \ \ \ 	\ \ \ \ \ 
		{(\textit{e})}: $n_a=201$, $\lambda=0$
          \ \ \ \ \ \ \
		{(\textit{f})}: $n_a=201$, $\lambda_{lim}=0.35911$
	\end{subfigure}
   \includegraphics[scale=0.525]{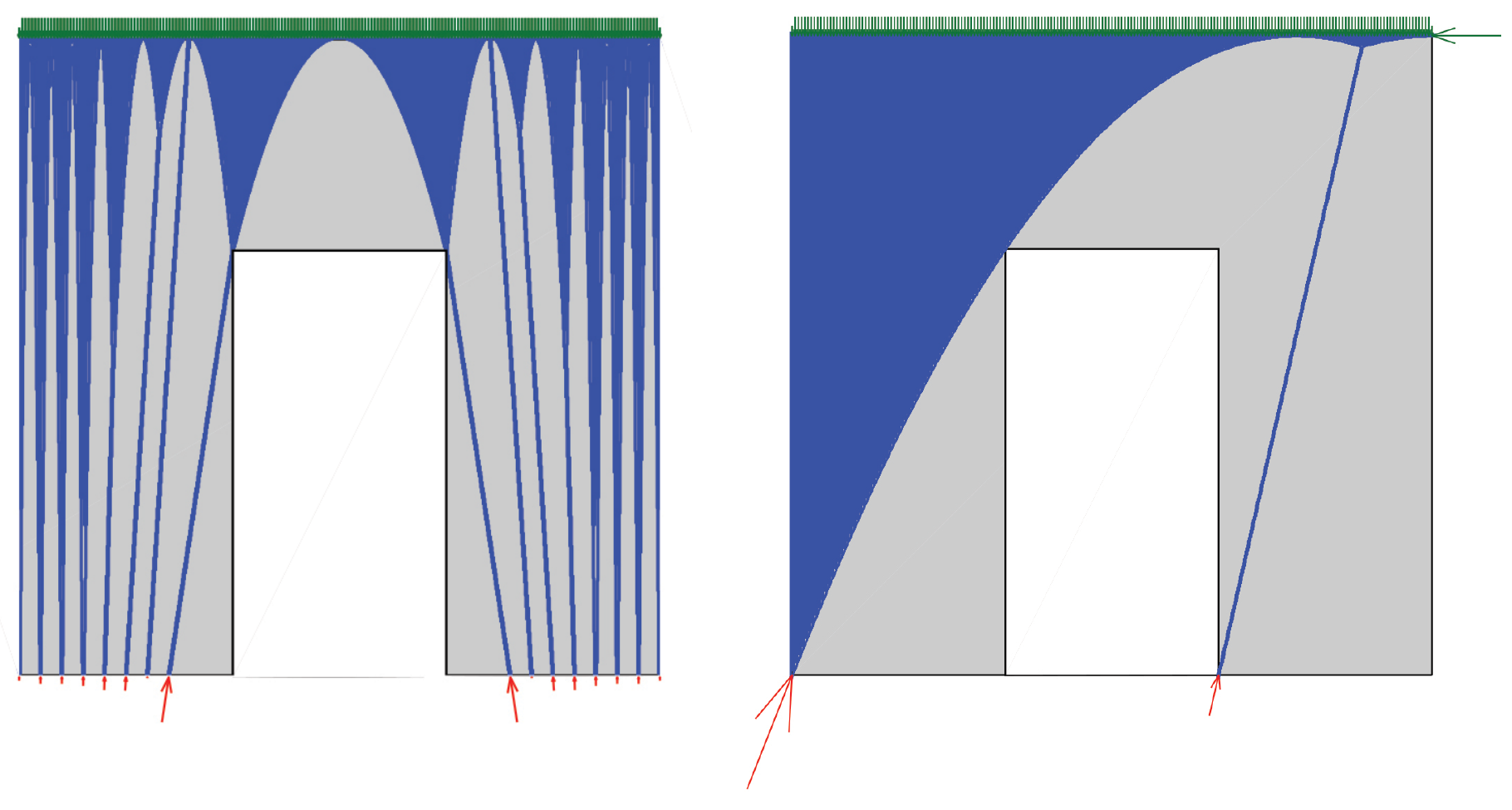} \\
    \caption{Strut net models of a masonry wall with one opening loaded by a uniformly distributed vertical load $q$ and concentrated horizontal force $F$.
     (\textit{a,b,c}): Loading conditions considering only the  distributed load $q$, which is lumped at $n_a=21,81$ and 201 points, respectively (the scaling factor of the vertical forces $s_q$ has been respectively set to 2, 8 and 20). The strut net models for the current loading conditions have been obtained via the LP procedure presented in \cite{obstacle22}.
     (\textit{b}),(\textit{d}),(\textit{e}): Loading conditions considering the vertical load $q$ and a concentrated horizontal force $F=\lambda q L$ applied to the top-right corner. Strut net models have been obtained through the LP procedure of section \ref{linearprogramming2D} (the scaling factor of $F$ has been assumed equal to one. Online version in color.)
    } 
    \label{wall_figure}
\end{figure*}

Figure \ref{wall_figure} shows strut net models of a one-story masonry wall with a central opening, which features overall length $L=2 b_1+ b_2=3$, overall height $H=h_1+h_2$ is 3, opening width $b_1=1$, and opening height $h_1=2$ equal to 2 (in abstract units).

\newpage

\begin{figure*}[h!]
   \centering	
\begin{subfigure}{0.675\textwidth}
   \includegraphics[scale=0.79]{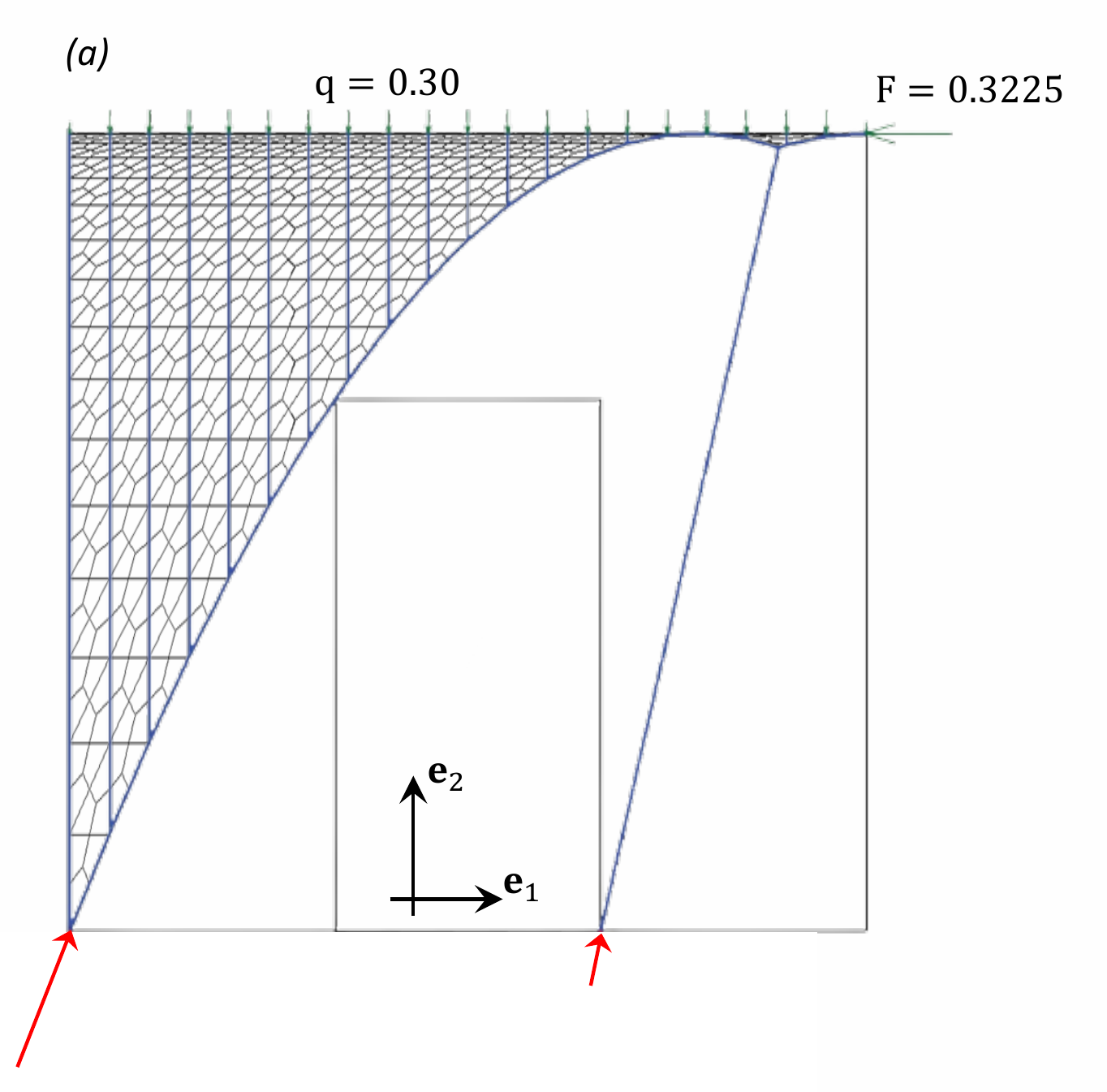}
\end{subfigure} \\
   \centering	
\begin{subfigure}{0.675\textwidth}
	\includegraphics[scale=0.79]{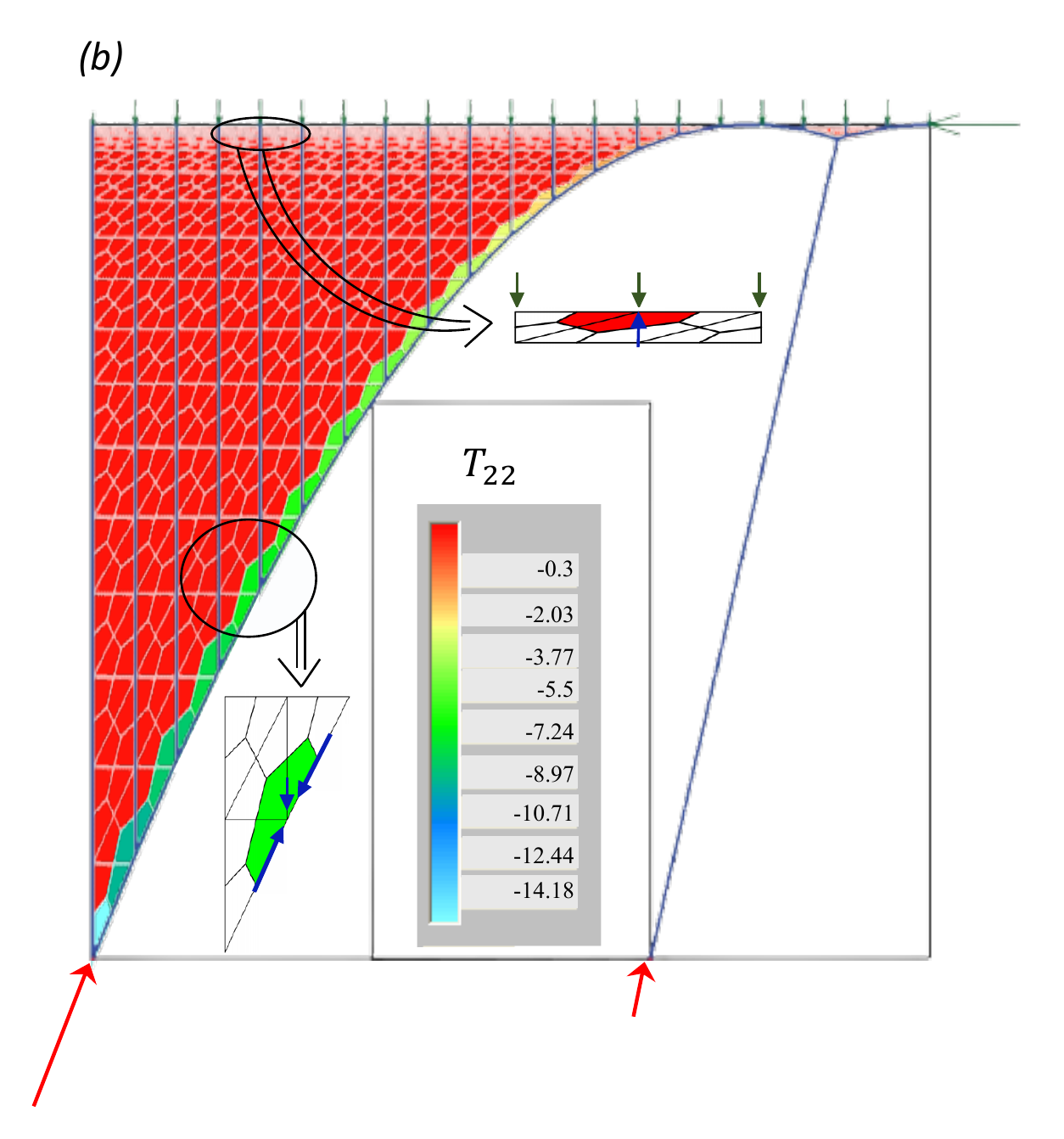}   
\end{subfigure}
     \caption{
 (\textit{a}) Primal and dual meshes of the region placed above the arched portion of the strut net shown in Fig. \ref{wall_figure}(\textit{b}).
 (\textit{b}) Density plot of the regularized stress component $T_{22}$. The inserts show two selected dual cells and the axial forces carried by the struts crossing such cells.
    (Online version in color.)
    }
         \label{T22map}
\end{figure*}

The wall is loaded with a uniform vertical load $q$ on the top edge, and a horizontal force $F=\lambda q L$ applied to the top-right corner and directed inside the wall.
Let us introduce the dimensionless variables $\xi=b_2/b_1=1$, and $\zeta=h_1/h_2=2$ as in \cite{silhavy2006}.
Reaction points were introduced at the base of the two piers placed on the sides of the opening, and the latter was modeled as an obstacle in the sense of the LP procedure illustrated in section \ref{linearprogramming2D}.
By lumping the ‘active’ load distribution $q$ in correspondence to a variable number of points $n_a$ uniformly spaced on the top edge, introducing $n_d=11$ uniformly spaced reaction points at the base of each pier, and using the above procedure, we obtained the limit load multipliers shown in figures \ref{wall_figure}(\textit{b,d,f}). Strut nets supporting the pure vertical loading condition ($\lambda=0$) are instead shown in figures \ref{wall_figure}(\textit{a,c,e}). The existence of such strut nets is in agreement with the compatibility condition $\zeta \le 4 \xi (1 +\xi)$ obtained in \cite{silhavy2006}.
\cb For growing values of $n_a$, one observes that the multipliers $\lambda_{lim}$ associated with the strut nets depicted in Figs. \ref{wall_figure}(\textit{b,d,f}) converge from below to the collapse multiplier $\lambda_c=0.35911$ obtained in \cite{silhavy2006} for the wall under examination.
This is not surprising, since the forces carried by such strut nets describe stress fields represented by measures that are in equilibrium with the given forces. One can average such \textit{measure stress fields}, using, e.g., the averaging procedure illustrated in \cite{Fraternali2002,amendola2020} or the integration procedure given in \cite{silhavy2014}, to obtain square integrable, no-tension stress fields that are in equilibrium with the loading condition $(q, F=\lambda qL)$ at the continuum level (\textit{regularized stress fields}). 
A regularization procedure of the singular stress field shown in figure \ref{wall_figure}(\textit{b}) is graphically illustrated in figure \ref{T22map} (assuming $q=0.3$ and $\lambda_{lim}=0.35833$, we applied a horizontal force $F=0.3225$ to the top-right corner of the wall).
The panel (\textit{a}) of such a figure shows two partitions of the region placed above the arched portion $\tilde{\cal S}$ of the overall strut net $\cal S$: a \textit{primal} triangulation, and \textit{dual mesh} formed by a centroidal Voronoi tessellation of the vertices of the primal mesh.
Figure \ref{T22map}(\textit{b}) illustrates the density plot of the $T_{22}$ component of the regularized stress field $\tilde{\BT}$, which is obtained making use of the lumped stress method presented in \cite{Fraternali2002}. The stress measures (or axial forces) carried by the struts of $\cal S$ are averaged over the 
polygonal cells of the dual mesh, through the procedure described in \cite{Fraternali2002,amendola2020}.
\crevvv ${\tilde T}_{22}=-q$ \cn in all the dual cells that do not intersect $\tilde{\cal S}$. Differently, 
one observes 
that the quantity  $|{\tilde T}_{22}|$ reaches values much greater than $q$ in correspondence to the dual cells that intersect $\tilde{\cal S}$. Such results are aligned with the continuous level solution of the wall problem under examination, which gives the stress field $\BT$ as the superimposition of a uniform stress field ${\BT}_r=-q  {\Be}_2 \otimes {\Be}_2$  ($\otimes$ denoting the dyadic product symbol) and a singular stress field $\BT_s$ concentrated on $\tilde{\cal S}$  \cite{silhavy2006}.
We conclude that the multipliers $\lambda_{lim}$ shown in Figs. \ref{wall_figure}(\textit{b,d,f}) are strongly compatible with the examined loading condition, and therefore provide lower bound estimates of $\lambda_c$ \cite{silhavy2014}. The results in Fig. \ref{wall_figure} clearly illustrate the convergence properties of the sequence of measure stress fields to the collapse stress field given in \cite{silhavy2006}.

\begin{figure*}[h!]
	\centering
\begin{subfigure}{0.40\textwidth}
		\centering
		{(\textit{a})}: $n_a=81$, $\lambda=0$
	\end{subfigure}
	\begin{subfigure}{0.40\textwidth}
		\centering
		{(\textit{b})}: $n_a=81$, $\lambda_{lim}=0.45$
	\end{subfigure}
   \includegraphics[scale=0.49]{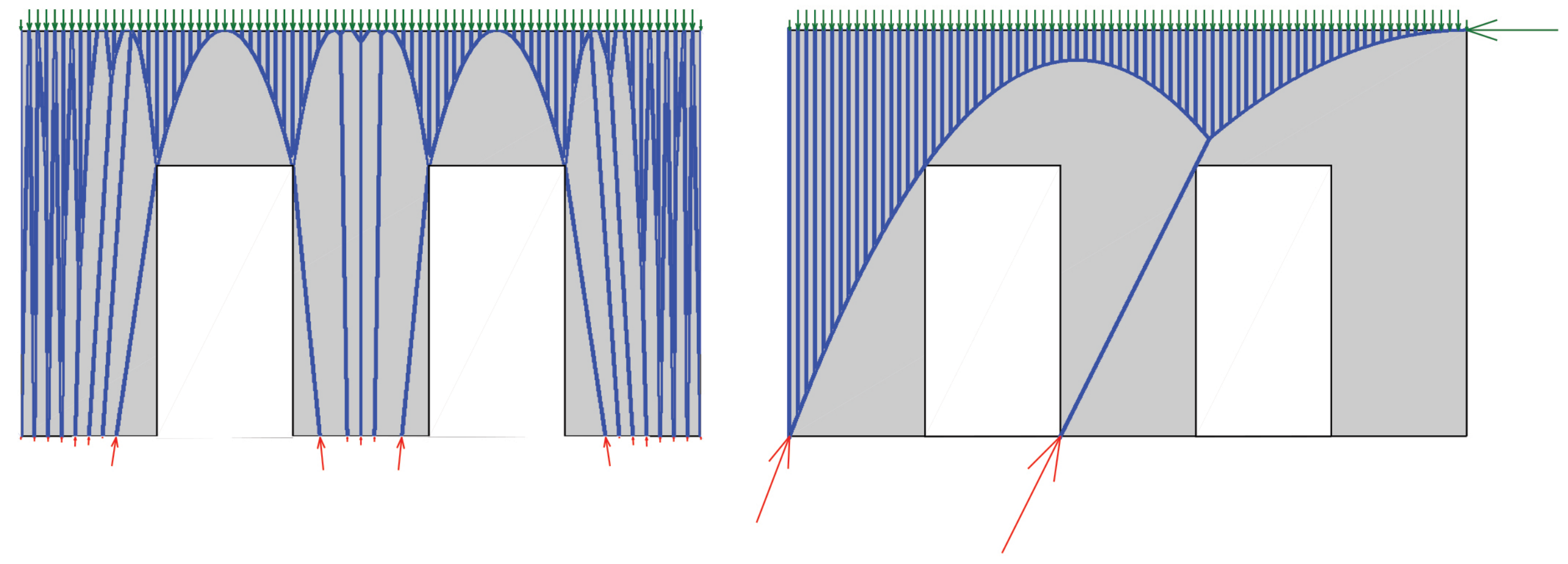} \\
   \begin{subfigure}{0.40\textwidth}
		\centering
		{(\textit{c})}: $n_a=11$, $\lambda_{lim}=0.1667$
	\end{subfigure}
	\begin{subfigure}{0.49\textwidth}
		\centering
		{\textit{b})}: $n_a=21$, $\lambda_{lim}=0.1667$
	\end{subfigure}
   \includegraphics[scale=0.49]{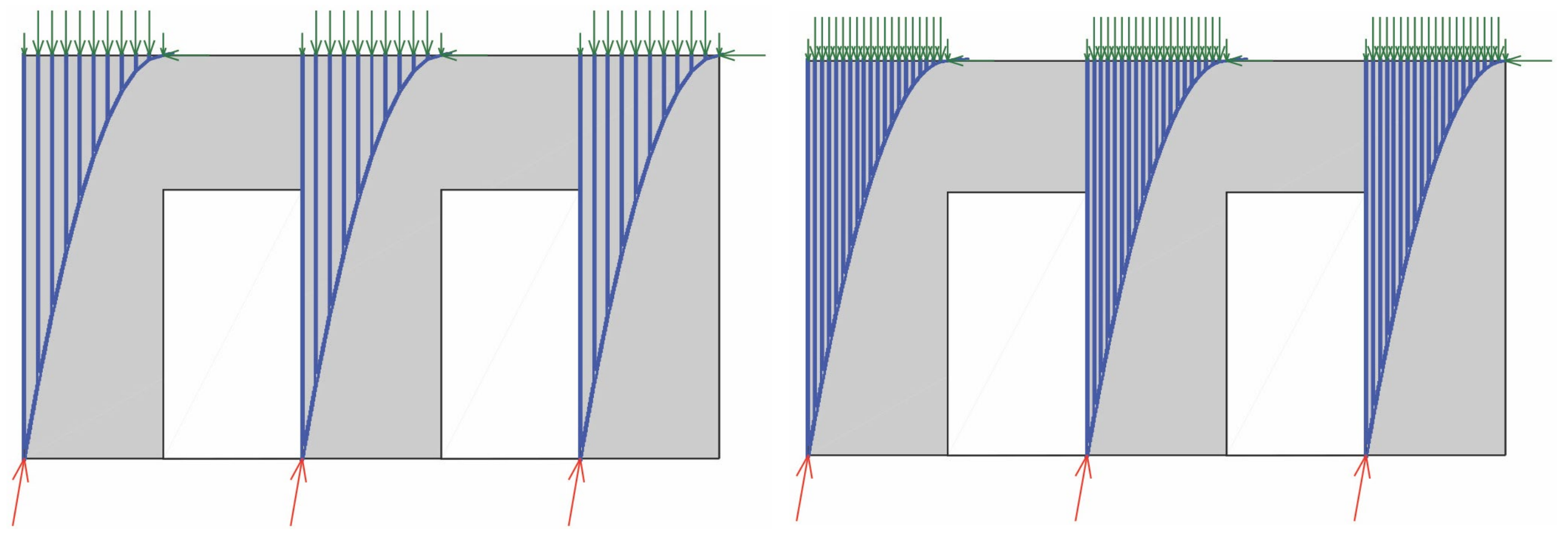} \\
\caption{Strut net models of a masonry wall with two openings loaded by uniformly distributed vertical loads $q$ and \crevvv a \cn concentrated horizontal force $F$.
     (\textit{a}) Case with the load $q$ distributed along the overall span $L$ of the top-edge. A strut net model is obtained through the LP procedure presented in \cite{obstacle22}, by lumping $q$ in correspondence to 81 points (the vertical active forces have been scaled by a factor $a_q=4$).
    (\textit{b}) Wall loaded by the same vertical forces as in panel (\textit{a}), and a concentrated horizontal force $F=\lambda q L$ in correspondence to the top-right edge of the wall. The limit load multiplier $\lambda_{lim}=0.45$ has been estimated making use of the LP procedure of section \ref{linearprogramming2D}. 
     (\textit{c}) Loading condition with two distributed loads $q$ acting on top of the piers and two horizontal forces $F=\lambda q b_1$ applied to the top-right corners of the piers. The vertical loads $q$ have been lumped in correspondence to $n_a=11$ points for each pier (\crevvv the vertical load $q$ and the horizontal force $F$ have been scaled by factors $s_q=6.67$ and $s_F=4$, respectively). \cn
    (\textit{d}) Loading condition similar to that analyzed in panel (\textit{c}), obtained by setting $n_a=21$ ($s_q=13.33$, $s_F=4$). %
    (Online version in color.)
    } 
    \label{frame_figure}
\end{figure*}

\cb
The successive figure \ref{frame_figure} shows strut net models of a masonry wall with two openings loaded by two different combinations of vertical and horizontal forces.
The overall length $L$ of the wall is 5; the height $H$ is 3, and the openings are uniformly spaced with width $b_1$ equal to 1 and height $h_1$ equal to 2 (as in the previous case).
Eleven reaction points were introduced at the bottom edges of each pier ($n_d=11$).
The example illustrated in figures \ref{frame_figure}(\textit{a,b}) examines a  loading condition similar to that examined in figure \ref{wall_figure}, which shows a uniformly distributed load $q$ applied along the entire span $L$ of the top edge of the wall, and a single horizontal force $F= \lambda qL$ applied to the top-right corner.
By setting $\lambda=0$; lumping the load $q$ in correspondence to $n_a=81$ points, and employing the LP procedure given in \cite{obstacle22}, we obtained the strut net shown in figure \ref{frame_figure}(\textit{a}), which proves that such a loading condition is supported by the wall.
Proceeding to examine the limit analysis problem in presence of the horizontal force $F= \lambda qL$, we employed the LP procedure of section \ref{linearprogramming2D} to obtain the limit load multiplier $\lambda_{lim}=0.45$, and the limit load strut net displayed in figure \ref{frame_figure}(\textit{b}).

\newpage

The example shown in figures \ref{frame_figure}(\textit{c,d}) refers to the wall loaded by three uniformly distributed loads with intensity $q$ applied on top of the piers, and by three horizontal forces $F= \lambda q b_1$ applied to right edges of the piers.
By lumping the ‘active’ load distribution $q$ in correspondence to $n_a=11$ points and using such a procedure, we obtained the limit load multiplier $\lambda_{lim}=0.1667$ and the limit load strut net displayed in figure \ref{frame_figure}\textit{a}.
An identical value of the limit load multiplier was obtained by setting $n_a=21$, as shown in figure \ref{frame_figure}\textit{b}. These results are in agreement with the collapse load multiplier $\lambda_c=1/6$ predicted by the kinematic approach to the limit analysis of masonry structures 
in correspondence to a ‘frame-type’ collapse mechanism (see, e.g., \cite{como2015}, and references therein).
The loading condition with $\lambda =0$ is supported by strut nets showing vertical struts running along the piers.

\cn

\subsection{Arched structure} \label{arch}

\begin{figure*}[h!]
	\centering
 	{(\textit{a})}: loading scheme \\  
\includegraphics[scale=0.25]{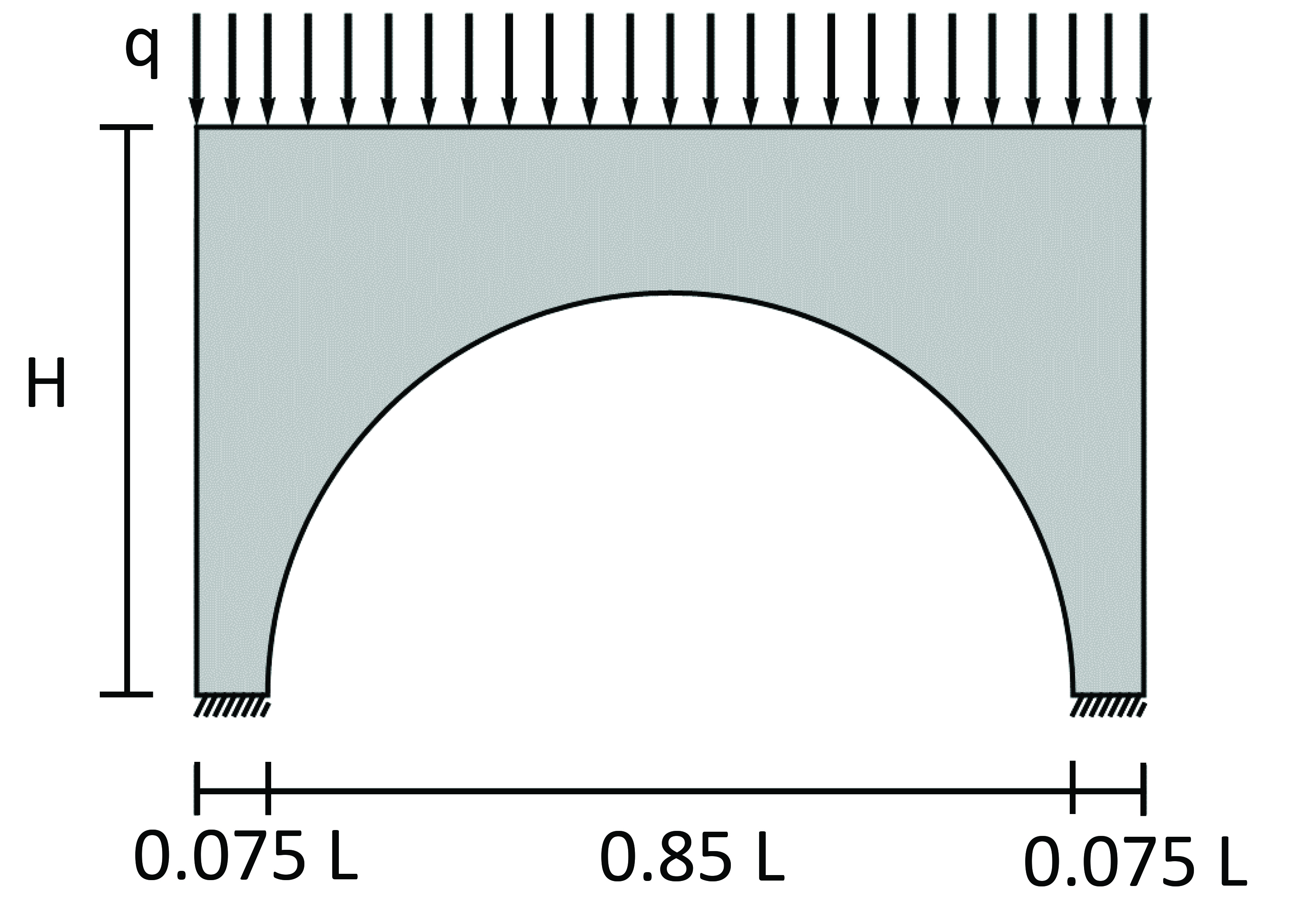} \\
\vskip 0,4cm
	\begin{subfigure}{0.40\textwidth}
	\centering
		{(\textit{b})}: $n_a=n_d=21$
		\includegraphics[width=\textwidth]{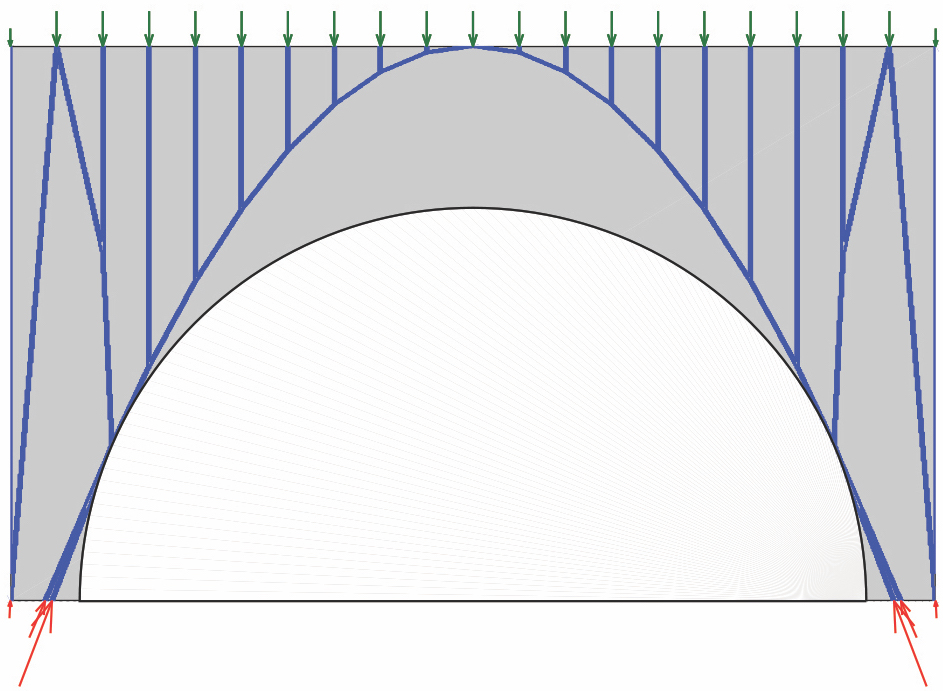}
\end{subfigure}
	\begin{subfigure}{0.40\textwidth}
		\centering
		{(\textit{c})}: $n_a=n_d=41$
		\includegraphics[width=\textwidth]{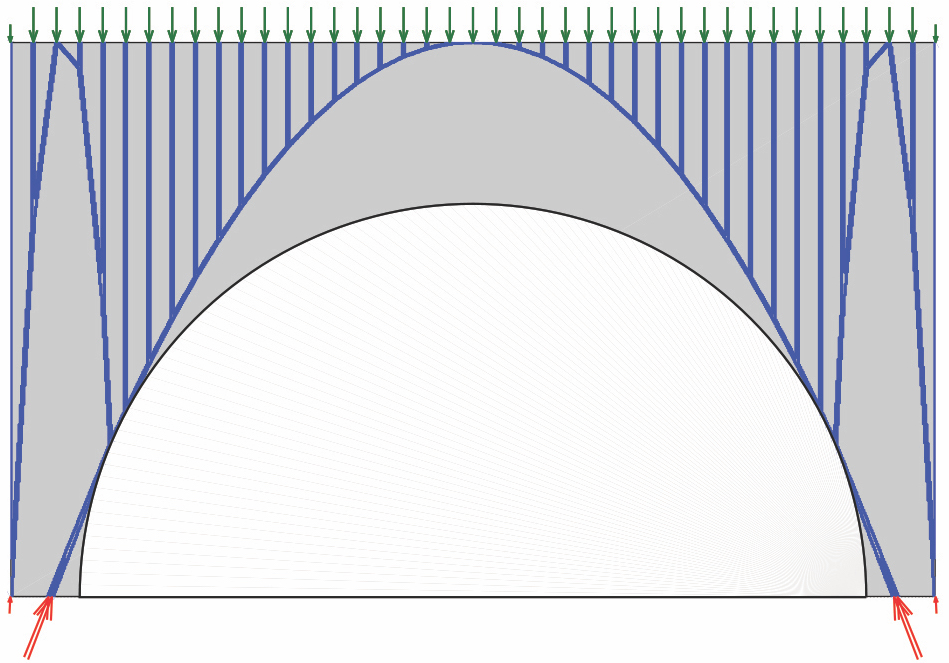}
	\end{subfigure}
\vskip 0,4cm	\begin{subfigure}{0.40\textwidth}
		\centering
		{(\textit{d})}: $n_a=n_d=101$
		\includegraphics[width=\textwidth]{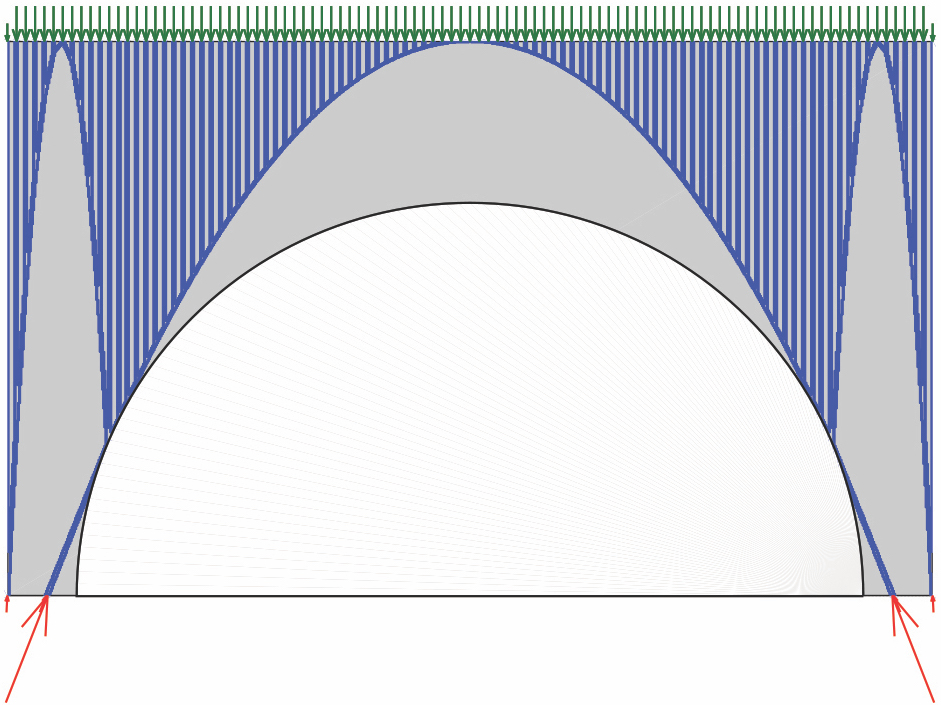}
	\end{subfigure}
	\begin{subfigure}{0.40\textwidth}
		\centering
		{(\textit{e})}: $n_a=n_d=201$
		\includegraphics[width=\textwidth]{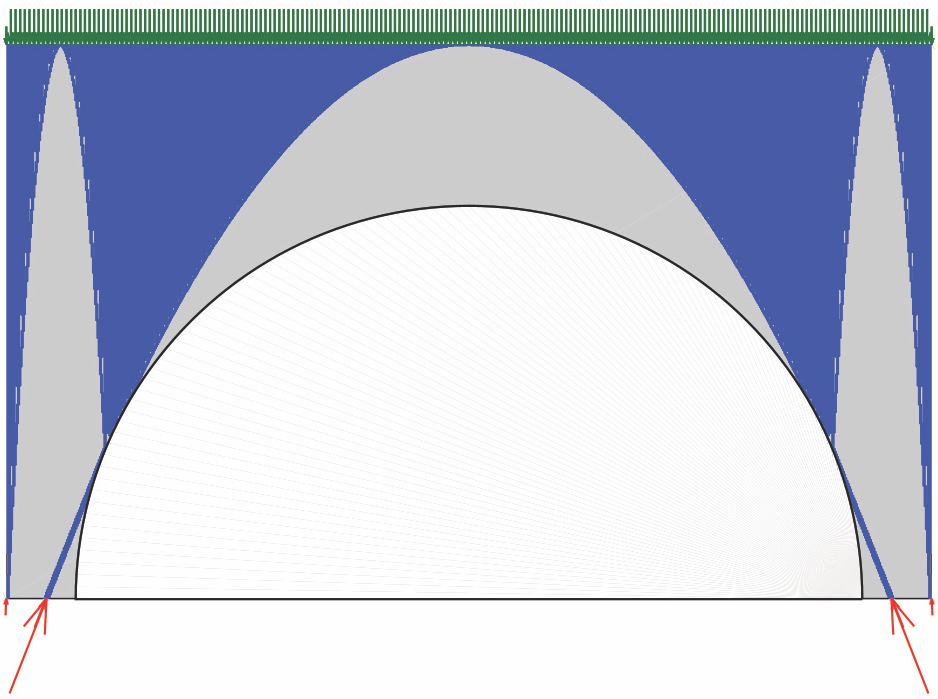}
	\end{subfigure}
    \caption{(\textit{a}) Masonry arch loaded by a uniform vertical load on the top edge.
    (\textit{b-e}) Strut net models obtained for $n_a=n_d=21$, 41, 101 and 201, respectively, by setting the scale factor $s_a$ of the vertical active forces acting on the top edge equal to 2.5, 5.0, 12.5, and 25. 
    (Online version in color.)}
    \label{arch_figure_1}
\end{figure*}

Figure \ref{arch_figure_1}\textit{a} shows a masonry arch that is loaded by a uniform vertical loading $q$ on the top edge.
We set $L/H=5/3$ and analyze different discretizations of the vertical load condition, which correspond to lumping $q$ at variable numbers $n_a$ of nodes uniformly spaced on the top edge. Each pier is constrained by $n_d$ (potential) reaction points at its base.
Figures \ref{arch_figure_1}\textit{b-e} show the strut nets that were obtained through the LP procedure given in \cite{obstacle22}, by letting $n_a=n_d$ vary from 21 to 201. It is seen that the arch supports the examined vertical loading condition.

\begin{figure*}[h!]
	\centering
 	{(a)}: loading scheme \\  
\includegraphics[scale=0.25]{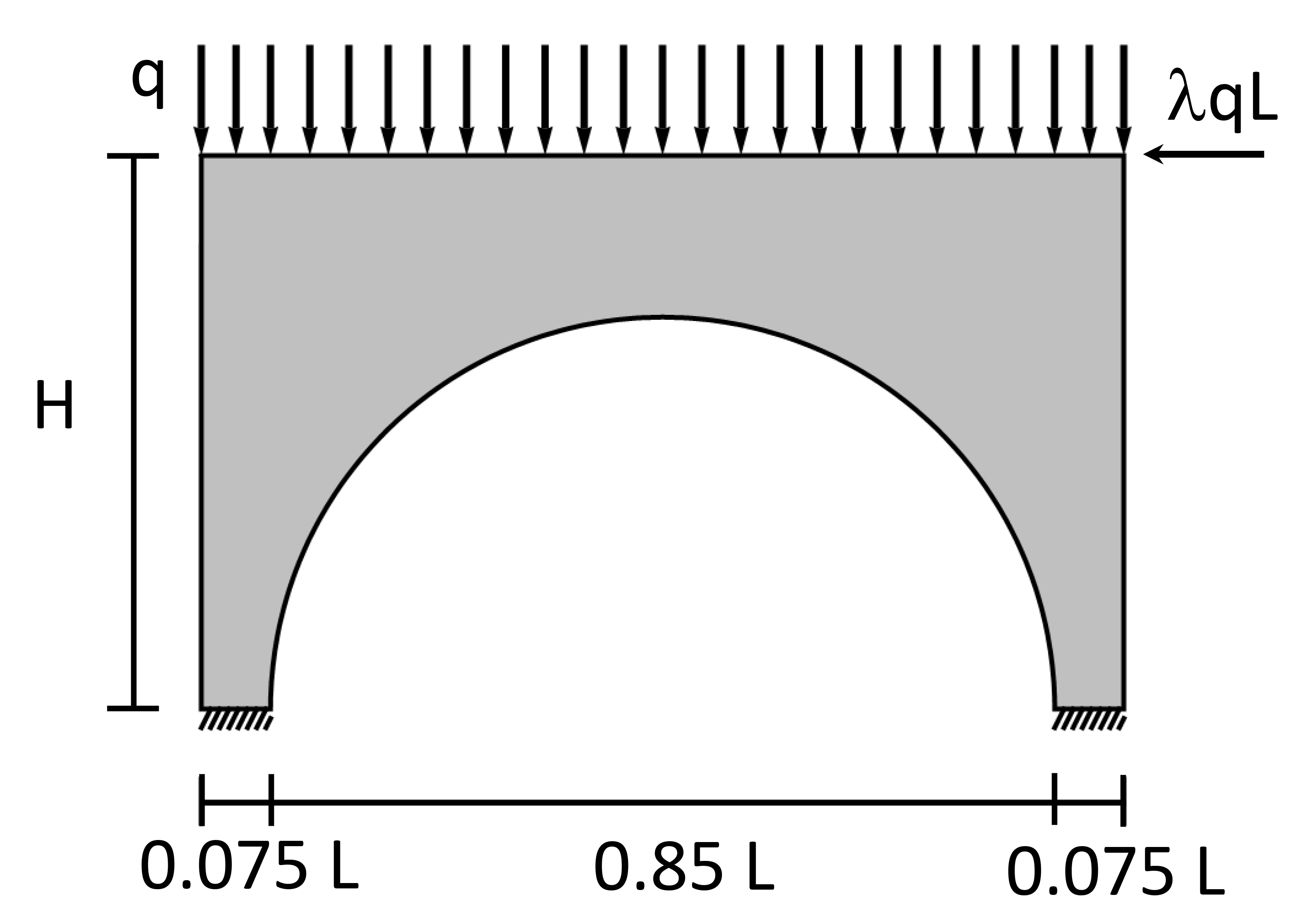} \\
\vskip 0,4cm
	\begin{subfigure}{0.49\textwidth}
	\centering
		{(b)}: $n_a=n_d=21$, 
		$\lambda_{lim}=0.3069$
		\includegraphics[width=\textwidth]{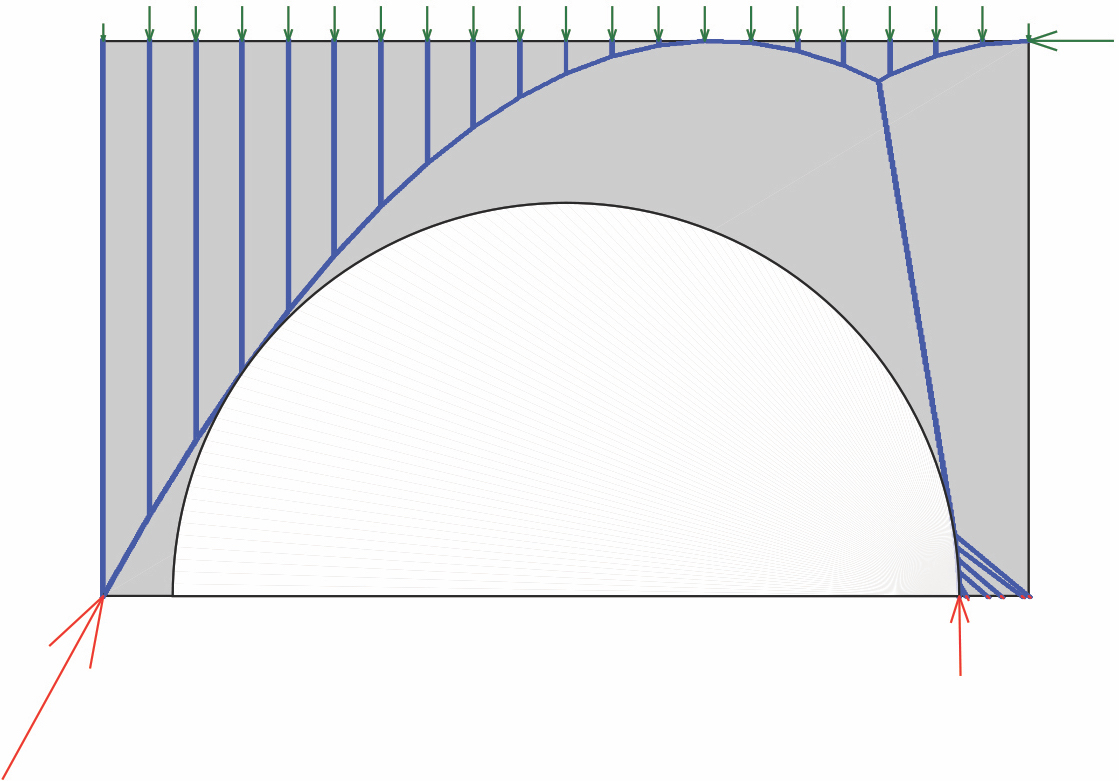}
\end{subfigure}
	\begin{subfigure}{0.49\textwidth}
		\centering
		{(c)}: $n_a=n_d=41$,
		$\lambda_{lim}=0.3080$
		\includegraphics[width=\textwidth]{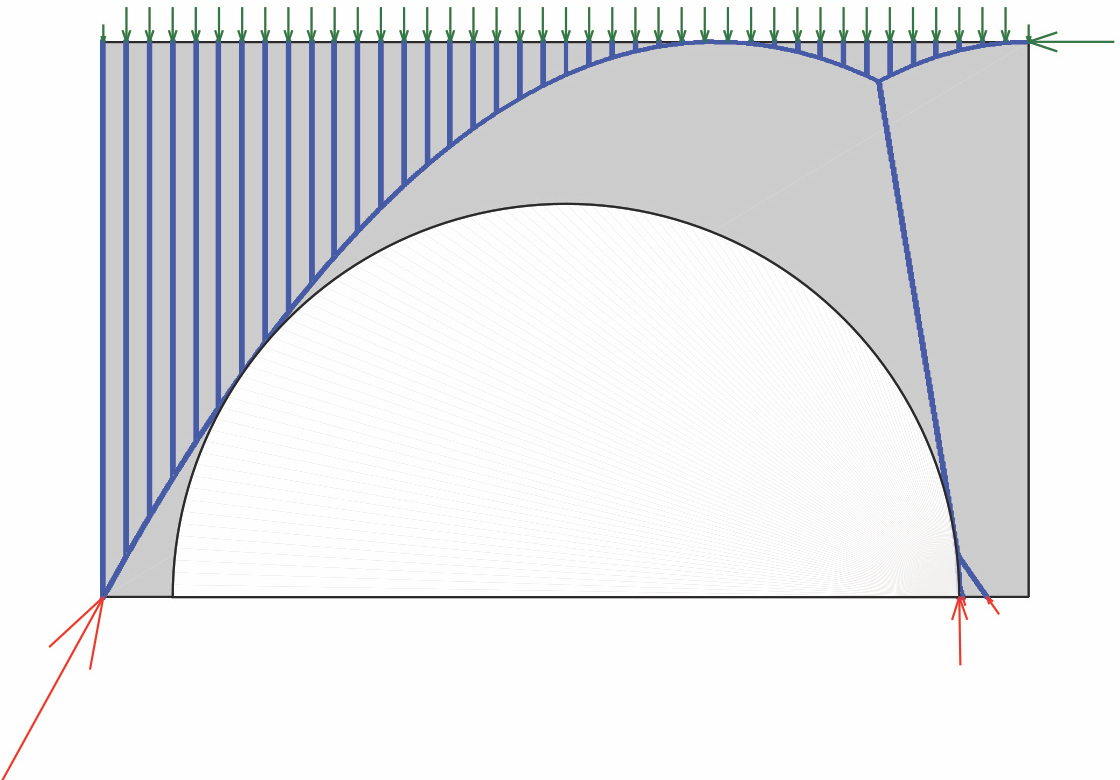}
	\end{subfigure}
\vskip 0,4cm
	\begin{subfigure}{0.49\textwidth}
		\centering
		{(d)}: $n_a=n_d=101$,
        $\lambda_{lim}=0.3083$
		\includegraphics[width=\textwidth]{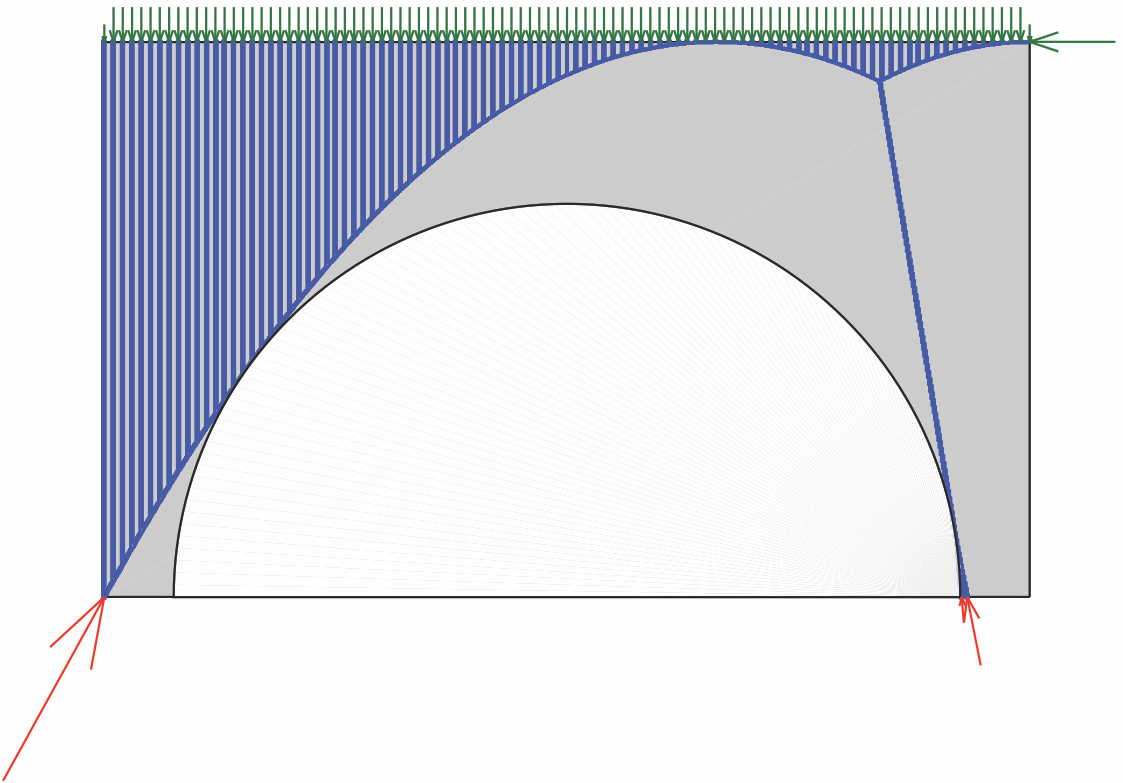}
	\end{subfigure}
	\begin{subfigure}{0.49\textwidth}
		\centering
		{(e)}: $n_a=n_d=201$, 
		$\lambda_{lim}=0.3084$
        \includegraphics[width=\textwidth]{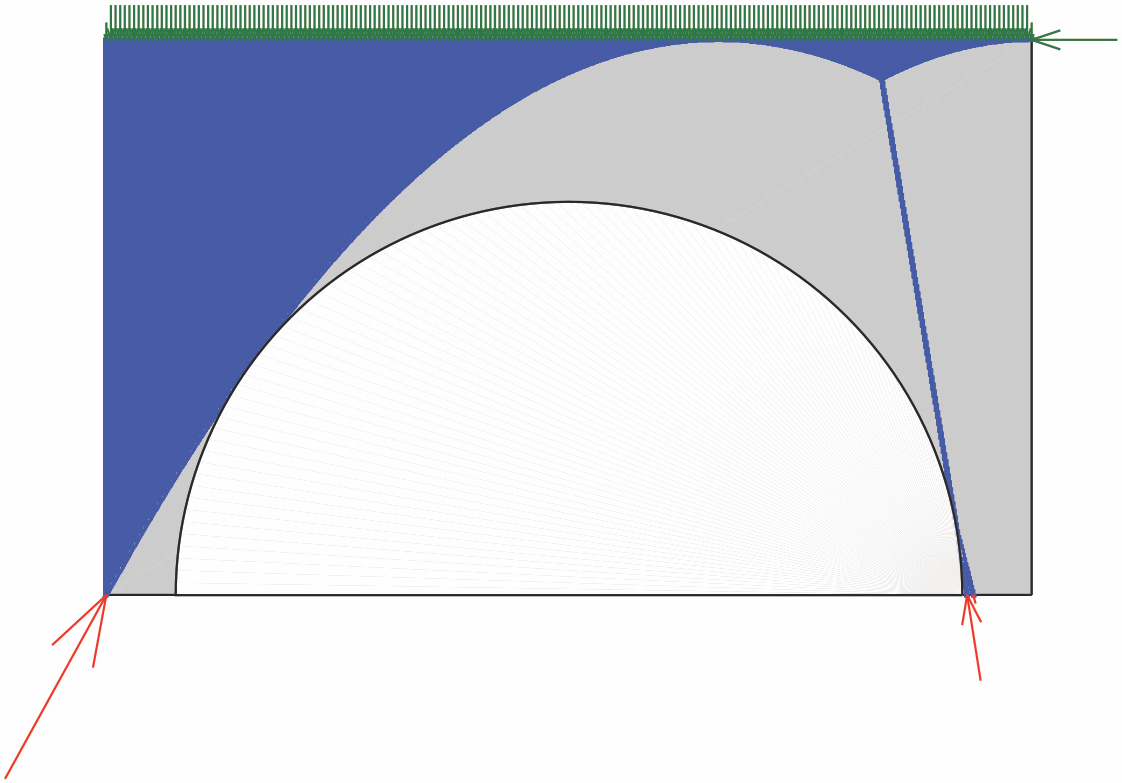}
	\end{subfigure}
	 \caption{(\textit{a}) Limit analysis problem of a masonry arch loaded by a uniform vertical load $q$ acting on the top edge and a concentrated horizontal force $F=\lambda qL$ applied to the top-right corner.
    (\textit{b-e}) Limit load strut nets obtained for $n_a=n_d=21$, 41, 101 and 201, respectively, by setting the scale factor $s_a$ of the vertical active forces acting on the top edge equal to 2.5, 5.0, 12.5 and 25. 
    (Online version in color.)}
    \label{arch_figure_2}
\end{figure*}

Figure \ref{arch_figure_2}\textit{a} illustrates the limit analysis problem obtained by applying a horizontal force $F=\lambda qL$ to the top-right corner of the arch.
We employed the LP procedure of section \ref{linearprogramming2D} to solve such a problem, by modeling the region underneath the arch as an obstacle.
The solutions obtained for the same values of $n_a=n_d$ analyzed for the vertical load condition are presented in figures \ref{arch_figure_2}\textit{b-e}.
One observes that the LP procedure returns estimates of the limit load multiplier that asymptotically converge from below to $\lambda_c=0.3084$. 
This is the collapse multiplier obtained through the continuum-level approach presented in section 3.3 of Ref. \cite{fortunato2013}, which is illustrated by Movie S2 in the supplementary materials.
\cb Making use of a regularization procedure similar to that illustrated in section \ref{walls}, it is possible to average the singular stress fields associated with the strut nets shown in figures \ref{arch_figure_2}(\textit{b-e}), so as to generate square integrable stress fields in equilibrium with the loading condition of figure \ref{arch_figure_2}(\textit{a}). \cn

\section{Concluding remarks} \label{conclusions}

\cb We have presented analytic and numerical results for the limit analysis problem of strut net discretization of masonry bodies, which are described through the rigid-no-tension constitutive model by Heyman \cite{heyman1995}.
Linear programming algorithms have been formulated to numerically solve problems dealing with simply- and multiply-connected domains. The predictions of such procedures have been validated against available results in the literature that benchmark examples of masonry walls and arches. 
Under suitable regularization conditions, the measure stress fields and the limit load multipliers predicted for strut net models of masonry structures can be strongly admissible for the underlying masonry structures, thus providing lower bounds of the collapse load multiplier $\lambda_c$ at the continuum level \cite{silhavy2014}.
A considerable part of the available results for the limit analysis of masonry structures is achieved through kinematic approaches (see, e.g., \cite{como2017} and references therein).
When the above regularization assumptions are matched,  the results presented in this study make it possible to complement the upper bound predictions of kinematic approaches with rigorous and easy-to-compute lower bounds of the collapse multipliers.
Having upper and lower bounds of $\lambda_c$ available, it is possible to predict such a quantity with sufficient accuracy, which is of paramount importance in technical applications \cite{como2017}.

We leave the issue of the 3D generalization of the obstacle problem presented in section  \ref{linearprogramming2D} to future work. 
We anticipate that such an extension, \crevvv
perhaps achieved \cn through the use of Beltrami stress functions or nonlinear mathematical programming procedures \cite{gill},
will not be straightforward and will pose complex analytic and numerical challenges.

\cn

\vskip6pt

\enlargethispage{20pt}

\noindent

\medskip


\begin{thebibliography}{00}


\bibitem{delpiero1998} Del Piero, G. Limit analysis and no-tension materials. \textit{Int. J. Plast} 1998;
\textbf{14}: 259-271. 

\cb

\bibitem{silhavy2014} 
\v{S}ilhav\'y, M. 2014. Mathematics of the Masonry–Like model and limit analysis. 
In: Angelillo, M. (ed.) Mechanics of Masonry Structures. CISM International Centre for Mechanical Sciences, vol 551. Springer, Vienna. 

\cn

\bibitem{miltonrspa2019} Bouchitt\'{e}, G., Mattei, O., Milton, G.W., Seppecher, P. On the forces that cable webs under tension can support and how to design cable webs to channel stresses. \textit{Proc. Math. Phys. Eng. Sci.} 2019; \textbf{475}: 2223. 

\bibitem{miltonSIAM2020} Bouchitt\'{e}, G., Mattei, O., Milton, G.W., Seppecher, P. Guiding stress with cable networks and the spider web problem. \textit{SIAM News} 2019; October 2020. 

\bibitem{heyman1995} Heyman, J. 1995  \textit{The Stone Skeleton: Structural Engineering of Masonry Architecture}.
Cambridge: Cambridge University Press.


\bibitem{Kamen1996}
Kamenjarzh, J. \textit{Limit Analysis of Solids and Structures}. Boca Raton: CRC Press 1996.

\bibitem{Save1997}
Save, M.A., Massonet, C.E., De Saxce, G. \textit{Plastic Limit Analysis of Plates, Shells and Disks} 1997. Amesterdam: Elsevier.

\bibitem{Drucker1952} Drucker, D.C., Prager, W., Greenberg, H.J.  Extended limit design theorems for continuous media.  \textit{Q. Appl. Math.} 1952; \textbf{9}: 381–389.
 
\bibitem{Giaquinta1985} Giaquinta, M., Giusti, E.  Researches on the equilibrium of masonry structures. \textit{Arch. Ration. Mech. Anal.} 1985; \textbf{88}: 359-392. 
 
\bibitem{AngelilloJomms2010} Angelillo, M., Cardamone, L., Fortunato, A. A Numerical model for masonry-like structures. \textit{J. Mech. Mater} 2010; \textbf{5}: 583-615. 

\bibitem{odwyer1999} O'Dwyer D. Funicular analysis of masonry vaults \textit{Comput. Struct.} 1999; \textbf{73}: 187-197. 

\bibitem{block2007} Block, P.,  Ochsendorf, J. Thrust network analysis: A new methodology for three-dimensional equilibrium \textit{J. Int. Assoc. Shell Spat. Struct.} 2007; \textbf{48}: 167-173. 

\bibitem{lucchesi2008} Lucchesi, M., Padovani, C., Pasquinelli, G., Zani, N. Masonry constructions: Mechanical models and numerical applications \textit{Lecture Notes in J. Appl. Comput. Mech.} 2008; \textbf{39}: 1-168. 

\bibitem{fortunato2010} Fortunato A.  Elastic solutions for masonry-like panels. \textit{J. Elast.} 2010; \textbf{98}: 87-110. 

\bibitem{defaveri2013} De Faveri, S., Freddi, L., Paroni, R.  No-tension bodies: A reinforcement problem. \textit{Eur. J. Mech., A/Solids} 2013; \textbf{39}: 163-169. 


\bibitem{como2017} Como, M. \textit{Statics of Historic Masonry constructions}, 2017.
3rd edn. Heidelberg: Springer.

\bibitem{Fraternali2002} Fraternali, F., Angelillo, M., Fortunato, A. A lumped stress method for plane elastic problems and the discrete-continuum approximation. \textit{Int. J. Solids Struct.} 2002; \textbf{39}: 6211–6240. 

\bibitem{fraternali2010} Fraternali, F. A thrust network approach to the equilibrium problem of unreinforced masonry vaults via polyhedral stress functions. \textit{Mech. Res. Commun.} 2010; \textbf{37}: 198-204. 

\bibitem{fortunato2013} Angelillo, M., Fortunato, A., Montanino, A., Lippiello, M. Singular stress fields in masonry structures: Derand was right. \textit{Meccanica} 2014; \textbf{49}: 1243-1262. 

\cred

\bibitem{milton2017} Milton, G.W. The set of forces that ideal trusses, or wire webs, under tension can support. \textit{Int. J. Solids Struct.} 2017; \textbf{128}: 272-281. 

\cn

\bibitem{pugno} Cranford, S.W., Tarakanova, A., Pugno, N.M., Buehler, M.J.  Nonlinear material behaviour of spider silk yields robust webs. \textit{Nature} 2012; \textbf{482}: 72-76. 

\bibitem{jiang} Jiang, Y., Nayeb-Hashemi, H.  Dynamic response of spider orb webs subject to prey impact. \textit{Int. J. Mech. Sci.} 2020; \textbf{186}: 105899. 






\bibitem{miniaci2016} Miniaci, M., Krushynska, A., Movchan, A.B., Bosia, F., Pugno, N.M. Spider web-inspired acoustic metamaterials. \textit{Appl. Phys. Lett.} 2016; \textbf{109}(7): 071905. doi:10.1063/1.4961307

\bibitem{jin2017}
Jin, Y., Yuan, H., Lan, J.L., Yu, Y., Lin, Y.H., Yang, X. Bio-inspired spider-web-like membranes with a hierarchical structure for high performance lithium/sodium ion battery electrodes: The case of 3D freestanding and binder-free bismuth/CNF anodes. \textit{Nanoscale} 2017; \textbf{9}(35): 13298-13304. doi:10.1039/c7nr04912a

\bibitem{cao2021}
Huang, H., Cao, E., Zhao, M., Alamri, S., Li, B. Spider web-inspired lightweight membrane-type acoustic metamaterials for broadband low-frequency sound isolation. \textit{Polymers} 2021; \textbf{13}(7): 1146. doi:10.3390/polym13071146

\cb

\bibitem{Zani2009}
Pintucchi, B., Zani, N. Effects of material and geometric non-linearities on the collapse load of masonry arches.  {\em Eur. J. Mech. A/Solids} 2009; \textbf{28}: 45-61.  



\bibitem{equivalent13}
Bourahla N. 2013. Equivalent Static Analysis of Structures Subjected to Seismic Actions. In: Beer M., Kougioumtzoglou I., Patelli E., Au IK. (eds) Encyclopedia of Earthquake Engineering. Springer, Berlin, Heidelberg.

\cn






\bibitem{2} Camar-Eddine, M., Seppecher, P. Determination of the closure of the set of elasticity functionals. \textit{Arch. Ration. Mechan. Anal.} 2003; \textbf{170}: 211-245.


\bibitem{5} Guevara Vasquez, F., Milton, G.W., Onofrei, D. Complete characterization and synthesis of the response function of elastodynamic networks. \textit{J. Elastic.} 2011; \textbf{102}: 31-54.


\cb

\bibitem{obstacle22} Amendola, A., Mattei, O., Milton, G.W., Seppecher, P. The obstacle problem in masonry structures and cable nets. \textit{arXiv:2204.00729v1} 2022.

\cn


\bibitem{MRCtensegrity} Skelton, R.E., Fraternali, F., Carpentieri, G., Micheletti, A. Minimum mass design of tensegrity bridges with parametric architecture and multiscale complexity. \textit{Mech. Res. Commun.} 2014; \textbf{58}: 124-132. 

\bibitem{SMStensegrity} Fraternali, F., De Chiara, E., Skelton, R.E. On the use of tensegrity structures for kinetic solar facades of smart buildings. \textit{Smart Mater. Struct.} 2015; \textbf{324}(10): 105032.


\bibitem{orduna17}
Ordu\~{n}a A. Non-linear static analysis of rigid block models for structural assessment of ancient masonry constructions,
\textit{Int. J. Solids Struct.} 2017;
\textbf{128}:
23-35.

\bibitem{lorenco05}
Louren\c{o}, P.B., Oliveira, D.V., Roca, P., Orduña, A. Dry joint stone masonry walls subjected to in-plane combined loading. \textit{J. Struct. Eng.} 2005; \textbf{131}(11): 1665-1673. 


\cb

\bibitem{silhavy2006} 
Lucchesi, M., \v{S}ilhav\'y, M., Zani, N. A new class of equilibrated stress fields for no-tension bodies. Journal of \textit{Mech. Adv. Mater. Struct.} 2006; \textbf{1}(3): 503-539. 

\bibitem{amendola2020} 
Amendola, A. On the optimal prediction of the stress field associated with discrete element models. \textit{J. Optim. Theory Appl.} 2020; \textbf{187}(3): 613-629.

\bibitem{como2015} 
Como, M. Seismic strength of reinforced multi-storey masonry walls. 
\textit{Key Eng. Mater.} 2015;
\textbf{624}, Trans Tech Publications, Ltd., Sept. 2014, 19–26.

\cn

\cred

\bibitem{gill}
Gill, P.E., Murray, W., Wright, M.H., 1981 \textit{Practical Optimization}. San Diego:  Academic Press, Inc.

\cn


\end{thebibliography}


\section*{Author contributions}
{All authors contributed equally and gave final approval for publication.}

\section*{Declaration of interests}
{The authors declare that they have no known competing financial interests or personal relationships that could have appeared to influence the work reported in this paper.}

\section*{Funding}
{OM and GWM are grateful to the National Science Foundation for support through Research Grants DMS-2008105 and DMS-2107926.
AA and FF are grateful to the Italian Ministry of University and Research  for support through the PRIN 2017 grant 2017J4EAYB.}


\section*{Data availability}

 Two movies are provided as supplementary material. Movie S1 illustrates the step-by-step loop reduction procedure for the example depicted in figure \ref{orduna_shear_figure}b-c, while Movie S2 illustrates the continuum-level solution for the limit analysis problem in figure \ref{arch_figure_2},

\section*{Supplementary materials}
Movie\_S1.mp4

\medskip

\noindent \textbf{Caption for Movie S1}

\medskip

\noindent A strut net describing the internal resisting structure of a shear wall made of dry stone masonry 
is simplified via the loop reduction procedure illustrated in \cite{miltonrspa2019}.

\medskip

\noindent Movie\_S2.mp4

\medskip

\noindent \textbf{Caption for Movie S2}

\medskip
\noindent Continuum-level solution for the limit analysis problem in figure \ref{arch_figure_2}, obtained through the procedure described in \cite{fortunato2013}.

\end{document}